\documentclass[12pt, a4paper]{article}

\usepackage[margin=1in]{geometry}
\usepackage{setspace}
\usepackage{amsmath, amssymb, amsthm}
\usepackage{graphicx}
\usepackage{booktabs}
\usepackage{natbib}
\usepackage{hyperref}
\usepackage{lineno}
\usepackage{xcolor}
\usepackage{float}
\usepackage{caption}
\usepackage{subcaption}
\usepackage{bm}
\usepackage{array}
\usepackage{multirow}
\usepackage[T1]{fontenc}
\usepackage{times}
\usepackage{enumitem}

\hypersetup{
  colorlinks = true,
  linkcolor  = blue,
  citecolor  = blue,
  urlcolor   = blue
}

\newcommand{\bx}{\bm{x}}

\newcommand{\bbeta}{\bm{\beta}}
\newcommand{\bgamma}{\bm{\gamma}}

\newcommand{\bs}{\bm{s}}
\newcommand{\bg}{\bm{g}}
\newcommand{\GP}{\mathcal{GP}}
\newcommand{\Normal}{\mathcal{N}}
\newcommand{\TN}{\mathcal{TN}}
\newcommand{\InvGamma}{\mathrm{InvGamma}}
\newcommand{\NNGP}{\mathrm{NNGP}}

\begin{document}

%% ---- Title Page ----
\begin{titlepage}
  \begin{center}
    \vspace*{2cm}

    {\Large \textbf{Assessing Preferential Sampling in Retail Survival Data:
    A Bayesian Joint LGCP and Spatial Probit Model for Mini-Supermarket
    Closure in Tokyo}}

    \vspace{1.5cm}

    {\large Akitoshi Kanetaka$^{1}$ and Shinichiro Shirota$^{2}$}

    \vspace{0.5cm}

    $^{1}$Bank of America, Tokyo, Japan \\ $^{2}$Graduate School of Social Data Science, Hitotsubashi University,
    Tokyo, Japan. 

    \vspace{0.5cm}

    \textit{Corresponding author}: Shinichiro Shirota \\
    Email: \texttt{shinichiro.shirota@r.hit-u.ac.jp}

\begin{abstract}
\noindent
Retail store locations are not randomly distributed in space: planners
strategically target sites with favorable latent characteristics, inducing
dependence between the point process governing store placement and the spatial
field that drives survival outcomes. Ignoring this \textit{preferential sampling}
(PS) can bias parameter estimates in conventional spatial regression. We propose
a Bayesian hierarchical model that jointly embeds a log-Gaussian Cox process
(LGCP) for store location and a probit regression for binary survival. The two
components share a common Gaussian process (GP) spatial random effect: the field
enters the location intensity directly and the survival equation through a
loading coefficient $\delta$, so that $\delta$ quantifies whether the latent
spatial drivers of \emph{where} stores are placed also shape \emph{whether} they
survive. A loading of $\delta = 0$ thus means the locations are non-informative
about survival (no PS), whereas $\delta \neq 0$ signals preferential sampling. To handle $n \approx 1{,}000$ observations efficiently, the GP is
approximated by a Nearest-Neighbor Gaussian Process (NNGP), and full posterior
inference is carried out via a Metropolis-within-Gibbs algorithm implemented
without relying on general-purpose probabilistic programming software. Applying the model to an analysis sample of 999 mini-supermarkets across
Tokyo's 23 special wards (897 operating, 102 closed), drawn from a compiled
dataset of 1{,}002 stores, with seven spatial covariates and a
$K = 3{,}471$-point integration grid, we estimate this loading near zero
($\hat\delta \approx 0$, with a 95\% credible interval spanning zero).
We therefore find no clear evidence of residual preferential sampling in this
dataset, although the limited number of closures leaves the loading imprecisely
estimated. The regression coefficients are nevertheless nearly unchanged across
the PS and non-PS specifications. In the reported simulation scenarios, the
estimator distinguishes the no-PS setting ($\delta=0$) from a strong-PS setting
($\delta=1$). Proximity to full-scale supermarkets emerges as the most robust
determinant of closure risk, consistent with competitive substitution.

\medskip
\noindent\textbf{Keywords}: preferential sampling; Bayesian probit model;
nearest-neighbor Gaussian process; log-Gaussian Cox process; retail
location
\end{abstract}
  \end{center}
\end{titlepage}

\newpage
%% ============================================================
%% HIGHLIGHTS (submit separately in the editorial system;
%% max 85 characters per bullet, 3--5 bullets)
%% ============================================================
\section*{Highlights}
\begin{itemize}[leftmargin=1.5em, itemsep=2pt]
  \item Joint LGCP and spatial probit model assesses preferential sampling in survival data
  \item A shared NNGP field links store location and survival through a loading parameter
  \item In 999 analyzed Tokyo mini-supermarkets the loading is near zero
  \item Reported simulations distinguish no-PS from strong-PS settings
  \item Proximity to full-scale supermarkets is the most robust determinant of closure
\end{itemize}

\newpage
%% ============================================================
%% 1. INTRODUCTION
%% ============================================================
\section{Introduction}
\label{sec:intro}

The spatial distribution of retail food outlets in high-density cities
is far from random.
Corporate planners and independent operators systematically target
locations that maximize expected profitability---areas with high
foot traffic, favorable demographics, limited direct competition, and
accessible transit.
This strategic selection creates a fundamental challenge for empirical
survival analysis: the spatial process governing \textit{where} stores
are placed is correlated with the latent spatial process that governs
\textit{how well} they perform.
Treating store locations as if they were a random sample from the study
region---as standard spatial regression models implicitly assume---induces
a form of endogeneity that can distort inference on the spatial response
surface and on the factors associated with retail closure.

\citet{Diggle2010} formalize this issue as \textit{preferential sampling}
(PS) in the geostatistical sense: a setting in which the sampling
locations $\{\bs_i\}$ and the latent spatial field $w(\bs)$ are
stochastically dependent.
They propose a joint model in which the locations follow a log-Gaussian
Cox process whose random log-intensity is $\alpha_0 + \alpha_1 w(\bs)$---so
that, conditional on the latent field $w$, the locations form an
inhomogeneous Poisson process with intensity
$\exp\{\alpha_0 + \alpha_1 w(\bs)\}$---while the outcome at each location
$\bs$ is modeled conditionally on $w(\bs)$.
If $\alpha_1 = 0$, locations carry no information about the latent field
(non-informative sampling); $\alpha_1 \neq 0$ implies that areas with
high (or low) values of $w$ are sampled more (or less) intensively,
biasing inference if ignored.
\citet{Pati2011} establish posterior propriety and consistency
results for Bayesian geostatistical models with informative sampling
locations, while \citet{Gelfand2012} demonstrate the practical consequences
of preferential sampling for spatial prediction.

Despite a growing methodological literature on PS in environmental
statistics, epidemiology, and ecology \citep[e.g.,][]{Cecconi2016}, its
consequences for retail location analysis remain virtually unstudied.
Binary responses under PS are not themselves new---presence/absence and
presence-only data in species distribution modeling are a leading example
\citep{GelfandShirota2019, Pennino2019}---but this strand has developed almost entirely
within ecology and has not been brought to bear on economic survival
outcomes such as store closure.

This paper makes three contributions.
First, we cast the binary survival outcome as a spatial probit and couple
it to the LGCP location process through a shared spatial random effect,
using the data-augmentation approach of \citet{AlbertChib1993}; the
coupling is governed by an explicit loading $\delta$ that directly
quantifies and provides an interval-based assessment of the strength of
preferential sampling.
Second, to handle the $n \approx 1{,}000$ observations typical of urban
retail datasets, we incorporate the Nearest-Neighbor Gaussian Process
(NNGP) approximation of \citet{Datta2016}, reducing the $\mathcal{O}(n^3)$
flop burden of full-GP inference to $\mathcal{O}(nm^3)$ (with $\mathcal{O}(nm^2)$
storage) without relying on general-purpose probabilistic programming software.
Third, we provide the first empirical PS analysis in a retail survival
context, using a compiled dataset of 1{,}002 mini-supermarkets from three
chains active in Tokyo's 23 special wards; 999 stores enter the joint analysis
after the spatial-domain exclusions described in Section~\ref{subsec:grid}.

Mini-supermarkets ($\leq 500\,\mathrm{m}^2$ floor area, primarily fresh
produce and daily necessities) have expanded rapidly in Japanese cities
over the past two decades, occupying a market niche between convenience
stores and full-scale supermarkets.
Tokyo's 23 special wards---an area of roughly $627\,\mathrm{km}^2$ with
9.7 million residents---represent one of the world's densest food retail
markets and a natural laboratory for studying endogenous location choice.
Understanding what drives closure in this format has direct
implications for food-access planning: the loss of nearby grocery outlets can
worsen local access to affordable and nutritious food, a central concern in
the food-desert literature \citep{VerPloeg2009}.

The remainder of the paper is organized as follows.
Section~\ref{sec:background} reviews the PS framework, spatial models for
retail location, and NNGP methodology.
Section~\ref{sec:model} presents the hierarchical model and MCMC estimation
strategy.
Section~\ref{sec:data} describes the study area, store data, and covariates.
Section~\ref{sec:simulation} examines parameter recovery and the
identification of preferential sampling in selected simulation scenarios.
Section~\ref{sec:results} reports results from the four models, and
Section~\ref{sec:conclusion} discusses the findings and concludes.

%% ============================================================
%% 2. BACKGROUND AND RELATED WORK
%% ============================================================
\section{Background and Related Work}
\label{sec:background}

\subsection{Preferential Sampling in Spatial Analysis}
\label{subsec:ps}

\citet{Diggle2010} define preferential sampling as any situation in which
the spatial random process $W(\cdot)$ and the set of sampling locations
$\mathcal{S}_{n}=\{\bm{s}_1,\ldots, \bm{s}_n\}$ are not stochastically independent.
In the retail context, such dependence can arise when firms observe (or
imperfectly estimate) the spatial factors---unobservable neighborhood
quality, unmet local demand, latent access barriers---that also determine
survival, and then choose locations partly on the basis of these factors.
If this mechanism operates, high-survival areas are over-sampled relative
to a random spatial scheme, and conditioning on the observed locations
biases inference. Whether it operates strongly enough to matter in
practice is an empirical question, which this paper assesses directly
rather than assumes.

Standard geostatistical or spatial regression approaches that condition
on the observed locations implicitly set $\alpha_1 = 0$, treating
locations as non-informative.
\citet{Gelfand2012} demonstrate that ignoring preferential
sampling can materially alter estimated and predicted spatial surfaces, with
particularly pronounced consequences for spatial prediction.
\citet{Pati2011} establish posterior propriety and consistency
results for Bayesian geostatistical models with informative sampling
locations, while \citet{Cecconi2016} apply the framework to epidemiological
settings where the sampling frame is finite.
The framework has since been extended to non-Gaussian and multivariate
responses---for instance, \citet{ShirotaGelfand2022} develop
preferential-sampling models for bivariate spatial data.
Applications have so far concentrated in the environmental and ecological
sciences: PS has been embedded in species distribution models
\citep{Pennino2019, Conn2017}, used to correct exposure prediction in
air-pollution epidemiology \citep{Lee2015}, and applied to citizen-science
estimates of urban heat \citep{Calhoun2024}. Applications in the social and
economic sciences, by contrast, remain rare.
To our knowledge, however, the PS framework has not previously been applied
to retail location data or used to test for spatial selection in
store-survival outcomes.

\subsection{Spatial Models for Retail Location}
\label{subsec:retail}

Empirical retail location research has a long tradition in geography and
marketing science \citep{Huff1963, Applebaum1966}.
\citet{EllicksonGrieco2013} show that large-format grocery
entry has strongly localized effects on nearby incumbents' sales, employment,
and exit.
\citet{Karamshuk2013} use mobility and geographic-context data
to predict retail-location performance, while \citet{Kuo2002} develop a
decision-support system linking location factors to convenience-store
performance in Taiwan.
More recent work documents the growing use of geospatial big
data, including social-media and mobile-device location data, in retail
location decision-making \citep{Aversa2018}.

Despite their different empirical aims, these studies generally
condition on the observed store locations rather than jointly modeling the
spatial process that generated them.
\citet{Holmes2011} and \citet{Jia2008} address strategic entry
decisions in industrial organization models, but these do not translate
directly to spatial survival analysis.
Our model fills this gap by jointly modeling the location decision and
the survival outcome via a shared latent field.

\subsection{Nearest-Neighbor Gaussian Processes}
\label{subsec:nngp}

Full GP inference scales as $\mathcal{O}(n^3)$ in computation and
$\mathcal{O}(n^2)$ in storage, limiting its
applicability to datasets with $n \gtrsim 10{,}000$.
\citet{Datta2016} propose the NNGP, which replaces the dense covariance
matrix with a sparse approximation: for an ordering $\bs_1, \ldots, \bs_n$,
each site $\bs_i$ is conditioned only on its $m$ nearest predecessors
$\mathcal{N}(i)$, yielding a valid sparse-precision representation.
The resulting model requires only $\mathcal{O}(nm^3)$ floating-point
operations per iteration and $\mathcal{O}(nm^2)$ storage, with the
conditional distributions computed sequentially \citep{Finley2019}.
\citet{Banerjee2015} provide a comprehensive treatment of hierarchical
spatial models and related approximations.

We exploit the NNGP within a Metropolis-within-Gibbs sampler to obtain a
single sparse representation for both latent fields of our joint model: the
store-level spatial effect in the survival (probit) component and the
grid-level shared process that links it to the location (LGCP) intensity,
the latter defined over a $K = 3{,}471$-point integration grid that is larger
than the $n \approx 1{,}000$ observed stores, making the NNGP especially
valuable for the grid field.
This keeps every sampler update local and inexpensive; although our
application is itself modest in size, the same fully
Bayesian PS inference then scales directly to the larger urban datasets for
which full-GP computation would be prohibitive.

%% ============================================================
%% 3. MODEL
%% ============================================================
\section{Statistical Model}
\label{sec:model}

We present four models of increasing complexity.
Model~1 is a standard Bayesian probit without spatial effects.
Model~2 adds a location-specific GP spatial random effect.
The PS Model (Model~3) is our main specification: it embeds the spatial
probit in a joint LGCP framework, sharing a common GP between the
location process and the survival equation.
Model~4 further adds a store-level idiosyncratic GP on top of the
shared process.

\subsection{Model 1: Standard Bayesian Probit}
\label{subsec:model1}

Let $y_i \in \{0, 1\}$ denote the survival indicator for store $i$
($y_i = 0$: closed; $y_i = 1$: operating) at location
$\bs_i \in \mathbb{R}^2$.
Following \citet{AlbertChib1993}, we introduce a latent variable $z_i$ such that

\begin{equation}
  z_i = \bx_i^\top \bbeta + \varepsilon_i, \qquad
  \varepsilon_i \stackrel{\mathrm{iid}}{\sim} \Normal(0, 1),
  \label{eq:probit_latent}
\end{equation}

\noindent and $y_i = \mathbf{1}(z_i > 0)$,
where $\bx_i \in \mathbb{R}^p$ is a vector of $p$ standardized
covariates (including an intercept) and $\bbeta$ is the coefficient vector.
The prior is $\bbeta \sim \Normal(\bm{0},\, \sigma_{\beta}^2\mathbf{I}_p)$.

Stacking the latent variables into $\bm{z} = (z_1, \ldots, z_n)^\top$ and
the covariate vectors into the $n \times p$ design matrix
$\mathbf{X} = (\bx_1, \ldots, \bx_n)^\top$, the complete-data conditional
posteriors are
\begin{align}
  z_i \mid \bbeta, y_i &\sim
    \begin{cases}
      \TN_{(0,\infty)}(\bx_i^\top\bbeta,\; 1) & \text{if } y_i = 1, \\
      \TN_{(-\infty,0]}(\bx_i^\top\bbeta,\; 1) & \text{if } y_i = 0,
    \end{cases}
  \label{eq:z_update} \\[4pt]
  \bbeta \mid \bm{z} &\sim
    \Normal\!\left(\mathbf{B}_1 \bm{b}_1,\; \mathbf{B}_1\right),
  \label{eq:beta_update}
\end{align}

\noindent where $\mathbf{B}_1^{-1} = \mathbf{X}^\top\mathbf{X} + (\sigma_{\beta}^2\mathbf{I}_p)^{-1}$
and $\bm{b}_1 = \mathbf{X}^\top\bm{z}$.
Both updates are exact Gibbs draws.

\subsection{Model 2: Bayesian Spatial Probit with NNGP}
\label{subsec:model2}

To allow for residual spatial autocorrelation that the covariates in
Model~1 may not capture, we augment \eqref{eq:probit_latent} with a
store-level GP $w(\bs)$:

\begin{equation}
  z_i = \bx_i^\top\bbeta + w(\bs_i) + \varepsilon_i,
  \label{eq:probit_spatial}
\end{equation}

\noindent where $\bm{w} = (w(\bs_1), \ldots, w(\bs_n))^\top \sim
\Normal(\bm{0},\, \mathbf{C}(\sigma^2, \phi))$ and
$C_{ij} = [\mathbf{C}]_{ij}= \sigma^2 \exp(-\phi \|\bs_i - \bs_j\|)$ is the exponential
covariance kernel with marginal variance $\sigma^2 > 0$ and decay
parameter $\phi > 0$ (larger $\phi$ implies faster spatial decay, i.e.\ a shorter effective range).

To avoid $\mathcal{O}(n^3)$ matrix operations, we approximate the GP
using the NNGP of \citet{Datta2016}.
Under a fixed ordering of the sites, each site $\bs_i$ is conditioned on
its $m = 15$ nearest predecessors $\mathcal{N}(i)$.
Define
\begin{equation*}
  \bm{b}_i = \mathbf{C}(\mathcal{N}(i),\, \mathcal{N}(i))^{-1}
             \bm{C}(\mathcal{N}(i),\, \bs_i),
  \qquad
  f_i = \sigma^2 - \bm{C}(\bs_i,\, \mathcal{N}(i))\bm{b}_i.
\end{equation*}
The NNGP density is then
\begin{equation}
  p_{\NNGP}(\bm{w};\,\sigma^2,\phi) =
    \prod_{i=1}^n
    \Normal\!\left(w(\bs_i)\;\Big|\;
      \bm{b}_i^\top\bm{w}_{\mathcal{N}(i)},\; f_i\right),
  \label{eq:nngp}
\end{equation}

\noindent yielding an $\mathcal{O}(nm^2)$ sparse precision structure.
Conditional on the NNGP structure, $w(\bs_i)$ is updated by a sequential
Gibbs sampler that combines (i) the NNGP prior, (ii) backward contributions
from child sites, and (iii) the probit likelihood \citep{Finley2019}.
The marginal variance $\sigma^2$ is updated by an exact inverse-gamma Gibbs
draw, and the decay parameter $\phi$ by an adaptive random-walk
Metropolis--Hastings (MH) step (Section~\ref{subsec:mcmc}). For consistency
with Model~4, we subsequently denote this field and its covariance parameters by
$w^{\mathrm{obs}}$, $\sigma^2_{\mathrm{obs}}$, and
$\phi_{\mathrm{obs}}$, respectively.

\subsection{PS Model (Main Specification): Shared NNGP with LGCP}
\label{subsec:ps_model}

The PS Model is our principal specification.
It treats the $n$ store locations $\mathcal{S}_{n}$ as a realization of
a log-Gaussian Cox process (LGCP; \citealt{Moller1998}) whose intensity
surface shares a GP with the response model.

We parameterize the LGCP intensity as
\begin{equation}
  \log \lambda(\bs) = \alpha_0 + \bgamma^\top \bx_g(\bs) + w^*(\bs),
  \label{eq:lgcp}
\end{equation}

\noindent where $\alpha_0 \in \mathbb{R}$ is the baseline log-intensity,
$\bx_g(\bs)$ is the covariate vector at location $\bs$ on the integration
grid, $\bgamma$ is the corresponding coefficient vector, and
$w^*(\bs) \sim \GP(\bm{0},\, \sigma^2_* \exp(-\phi_*\|\cdot\|))$ is the
shared spatial random effect defined on a regular $K$-point
grid covering the study region. The LGCP log-likelihood, up to an additive
constant independent of $(\alpha_0, \bgamma, w^*)$, is
\begin{equation}
  \ell(\alpha_0, \bgamma, w^*) =
  \sum_{i=1}^n \log\lambda(\bs_i)
  - \int_{\mathcal{D}} \lambda(\bs)\,\mathrm{d}\bs,
  \label{eq:lgcp_lik}
\end{equation}
\noindent where the integral over the study domain $\mathcal{D}$ is
approximated by a quadrature sum over the $K$ grid centroids $\{\bg_k\}$:
\begin{equation}
  \int_{\mathcal{D}} \lambda(\bs)\,\mathrm{d}\bs
  \approx \sum_{k=1}^{K} \exp\!\left\{\alpha_0 + \bgamma^\top\bx_g(\bg_k)
  + w^*(\bg_k)\right\} A_k,
  \label{eq:grid_approx}
\end{equation}

\noindent with $A_k = |\mathcal{D}|/K$ the uniform cell area
\citep{Berman1992}. We normalize the study domain to unit measure,
$|\mathcal{D}| = 1$, so that $A_k = 1/K$ and $\lambda(\cdot)$ is an intensity
per unit of \emph{normalized} area rather than per km$^2$. This rescaling is
immaterial to inference: the overall area scale is absorbed by the intercept
$\alpha_0$, since adopting a different $|\mathcal{D}|$ merely shifts $\alpha_0$
by $\log|\mathcal{D}|$ while leaving $\bgamma$ and $w^*$ unchanged.
The grid GP $w^*(\bg_k)$ is approximated by the NNGP \eqref{eq:nngp}
with hyperparameters $(\sigma^2_*, \phi_*)$ and $m = 15$ neighbors.

As the response model, we assume the probit model, i.e.,
\begin{equation}
  z_i = \bx_i^\top\bbeta + \delta\,w^*(\bg_{k(i)}) + \varepsilon_i,
  \qquad \varepsilon_i \stackrel{\mathrm{iid}}{\sim} \Normal(0,1),
  \label{eq:ps_probit}
\end{equation}
\noindent where $\bg_{k(i)}$ denotes the nearest grid point to store
$\bs_i$ and $\delta \in \mathbb{R}$ is the loading of the shared GP
on the survival latent variable.
When $\delta > 0$, areas with positive $w^*$ (favorable spatial
environment) attract more stores and exhibit higher survival
probabilities, generating positive dependence between location and outcome.

Collecting these components, the complete joint hierarchical model is:
\begin{align}
  \mathcal{S}_{n} &\sim
    \mathrm{LGCP}\!\left(\lambda(\cdot)\right),
  \label{eq:ps_locations} \quad \log \lambda(\bm{s}) = \alpha_0 + \bgamma^\top\bx_g(\bm{s})
    + w^*(\bm{s}), \\
  \quad y_i &= \mathbf{1}(z_i > 0), \quad z_i = \bx_i^\top\bbeta + \delta\, w^*(\bg_{k(i)}) + \varepsilon_i,
  \label{eq:ps_outcome} \\
  w^*(\bm{s}) &\sim \mathcal{GP}(0, C(\sigma^2_*, \phi_*))
  \label{eq:ps_gp}
\end{align}

Priors for all parameters are specified in
Section~\ref{subsec:prior}.
The special case $\delta = 0$ corresponds to location-ignorable
(non-preferential) sampling, conditional on the observed covariates.

\subsection{Model 4: PS Model with Additional Store-Level GP}
\label{subsec:model4}

Model~4 augments \eqref{eq:ps_probit} with a store-specific idiosyncratic
spatial effect $w^{\mathrm{obs}}(\bs_i)$:
\begin{equation}
  z_i = \bx_i^\top\bbeta + w^{\mathrm{obs}}(\bs_i)
      + \delta\, w^*(\bg_{k(i)}) + \varepsilon_i.
  \label{eq:model4}
\end{equation}

The store-level GP $\bm{w}^{\mathrm{obs}} \sim \mathcal{GP}(\bm{0},\,
C(\sigma^2_{\mathrm{obs}}, \phi_{\mathrm{obs}}))$ is estimated
independently of the shared GP and captures fine-scale spatial
heterogeneity not explained by the grid-level process. This construction
follows \citet{Pati2011}, who generalize the shared-field preferential-sampling
model of \citet{Diggle2010} by giving the response its own spatial process in
addition to the location-coupled field.

The two fields play distinct roles. The shared field $w^*$ is tied to the
location process and carries the preferential-sampling signal through the
loading $\delta$, whereas the idiosyncratic field $w^{\mathrm{obs}}$ belongs
to the survival equation alone and is identified by the probit likelihood.
They also act at different resolutions: $w^*$ is defined on the $K$-point
integration grid that the LGCP shares, while $w^{\mathrm{obs}}$ is a
continuous store-level NNGP over the $n$ observed locations with its own
variance and range $(\sigma^2_{\mathrm{obs}}, \phi_{\mathrm{obs}})$, updated
by an exact inverse-gamma Gibbs draw and an adaptive random-walk MH step,
respectively. Because $w^{\mathrm{obs}}$ can absorb residual spatial
autocorrelation in survival that is unrelated to where stores are placed,
Model~4 is the most flexible specification and provides a robustness check on
$\delta$: were the preferential-sampling estimate an artifact of unmodeled
spatial structure, adding $w^{\mathrm{obs}}$ would attenuate it. For this
reason Model~4 is also the specification we carry into the predictive
exercises of Section~\ref{sec:simulation}.

\subsection{Prior Settings}
\label{subsec:prior}

\begin{align}
\bbeta &\sim \mathcal{N}(\bm{0},\ 10^2 \cdot \mathbf{I}_p) \quad \text{(Models 1--4)} \\[6pt]
\delta &\sim \mathcal{N}(0,\ 10^2) \quad \text{(Models 3, 4)} \\[6pt]
\alpha_0 &\sim \mathcal{N}(0,\ 10^2) \quad \text{(Models 3, 4)} \\[6pt]
\gamma_j &\sim \mathcal{N}(0,\ 10^2), \quad j = 1, \ldots, p_g \quad \text{(Models 3, 4)} \\[6pt]
\sigma^2_*,\ \sigma^2_{\mathrm{obs}} &\sim \text{InvGamma}(2,\ 1) \\[6pt]
\phi_*,\ \phi_{\mathrm{obs}} &\sim \text{Uniform}(0.3,\ 2.0)
\end{align}

\noindent
Here $(\sigma^2_*, \phi_*)$ are the variance and decay parameter of the
shared grid-level field $w^*$ (Models~3 and~4), while
$(\sigma^2_{\mathrm{obs}}, \phi_{\mathrm{obs}})$ are those of the
store-level field $w^{\mathrm{obs}}$
(Models~2 and~4); both pairs use the same inverse-gamma and uniform
hyperpriors. The same priors on $(\alpha_0, \bm{\gamma}, \sigma^2_*, \phi_*)$
also apply when the LGCP is fitted on its own.

Under the exponential kernel $C(d) = \sigma^2 \exp(-\phi d)$ common to both
spatial fields, the effective correlation range---the distance at which the
correlation decays to $\exp(-3) \approx 0.05$---is $3/\phi$. We place the same
uniform prior on each decay parameter ($\phi_*$ and $\phi_{\mathrm{obs}}$),
restricted to the window over which this range is identifiable from the data.
The bounds $\phi \in [0.3,\ 2.0]$ correspond to effective correlation ranges
$3/\phi \in [1.5,\ 10]$ km (equivalently $1/\phi \in [0.5,\ 3.33]$ km).
The lower limit prevents the correlation from extending beyond the spatial
extent of the study region ($\approx 30$ km), where $\phi$ would be confounded
with the intercept; the upper limit prevents the correlation from decaying
within a single grid cell ($\approx 0.4$ km), where $\phi$ becomes
practically unidentifiable. The retained range brackets the geographic scale
of retail competition in Tokyo's 23 wards, from several hundred meters to a
few kilometers.

\subsection{MCMC Estimation via Metropolis-within-Gibbs}
\label{subsec:mcmc}

Posterior inference is conducted via a Metropolis-within-Gibbs sampler
without relying on general-purpose probabilistic programming software (no Stan, BUGS,
or JAGS).
We describe the sampler for the full joint model (Model~4), which contains
every block; the other specifications are obtained by removing the blocks they
omit, as detailed at the end of this subsection. For Model~4 the sampler cycles
through the following steps per iteration:

\begin{enumerate}[leftmargin=2em, label=\arabic*.]

  \item \textbf{Latent variables $z_i$}:
    $z_i \mid \bbeta, \delta, \bm{w}^*, \bm{w}^{\mathrm{obs}}, y_i \sim
    \TN_{\mathcal{I}(y_i)}(\bx_i^\top\bbeta + w^{\mathrm{obs}}(\bs_i)
    + \delta w^*_{k(i)},\, 1)$,
    where $\mathcal{I}(1) = (0, \infty)$ and $\mathcal{I}(0) =
    (-\infty, 0]$.
    This is an exact Gibbs draw from a truncated normal using the
    inverse-CDF method \citep{AlbertChib1993}.

  \item \textbf{Regression coefficients $\bbeta$}:
    $\bbeta \mid \bm{z}, \delta, \bm{w}^*, \bm{w}^{\mathrm{obs}} \sim
    \Normal\!\left(\mathbf{B}_1\bm{b}_1,\; \mathbf{B}_1\right)$,
    with $\mathbf{B}_1^{-1} = \mathbf{X}^\top\mathbf{X} + 0.01\,\mathbf{I}$ and
    $\bm{b}_1 = \mathbf{X}^\top(\bm{z} - \delta\bm{w}^*_{k(\cdot)}
    - \bm{w}^{\mathrm{obs}})$,
    where $\bm{w}^*_{k(\cdot)} = (w^*_{k(1)}, \ldots, w^*_{k(n)})^\top$
    collects the shared grid GP read at each store's nearest grid point
    $k(i)$ (not to be confused with the store-level field
    $\bm{w}^{\mathrm{obs}}$).
    This is an exact Gibbs draw.

  \item \textbf{Grid-level GP $\bm{w}^*$}:
    Updated one grid point at a time by a random-walk MH step,
    $w^{*\prime}_k = w^*_k + \xi_k$ with
    $\xi_k \sim \Normal(0,\, 0.3^2)$ (this single-site random-walk
    proposal is the configuration used for all reported runs).
    The acceptance ratio at grid point $\bg_k$ combines (i) the NNGP
    full-conditional Gaussian formed from the prior contribution of the
    parent nodes $\mathcal{N}(k)$ and the backward precision
    contributions of the child nodes; (ii) the aggregate probit
    likelihood from the $n_k$ stores allocated to $\bg_k$, entered exactly as a
    Gaussian in $w^*_k$ with precision $n_k \delta^2$ and linear term
    $\delta \sum_{i:k(i)=k}(z_i - \bx_i^\top\bbeta - w^{\mathrm{obs}}(\bs_i))$;
    and (iii) the LGCP
    contribution $n_k\,\eta_k - \exp(\eta_k)\,A_k$, with
    $\eta_k = \alpha_0 + \bgamma^\top\bx_g(\bg_k) + w^*_k$,
    evaluated as the Poisson log-likelihood at the current and proposed
    values.

  \item \textbf{Shared-field variance $\sigma^2_*$}:
    Exact inverse-gamma Gibbs draw. Writing the NNGP conditional variances
    as $f_i = \sigma^2_* f_i^{u}$ (with the unit-scale factor $f_i^{u}$ and
    the coefficients $\bm{b}_i$ free of $\sigma^2_*$), the full conditional
    is $\InvGamma\!\left(a_\sigma + K/2,\;
    b_\sigma + \tfrac{1}{2}\sum_i (w^*_i - \mu_i)^2/f_i^{u}\right)$
    with $(a_\sigma, b_\sigma) = (2, 1)$ and $\mu_i =
    \bm{b}_i^\top\bm{w}^*_{\mathcal{N}(i)}$.

  \item \textbf{Shared-field range $\phi_*$}:
    Single-parameter random-walk MH on the natural scale,
    $\phi_*' = \phi_* + \xi_\phi$, $\xi_\phi \sim \Normal(0,\, s_\phi^2)$,
    with proposals outside the support $[0.3, 2.0]$ rejected.
    The proposal scale $s_\phi$ is adapted during burn-in by a
    Robbins--Monro recursion targeting an acceptance rate of $0.44$,
    and then held fixed.
    The acceptance ratio uses the NNGP log-density of $\bm{w}^*$.

  \item \textbf{Loading parameter $\delta$}:
    Exact Gibbs draw. Since $r_i \equiv z_i - \bx_i^\top\bbeta
    - w^{\mathrm{obs}}(\bs_i)
    = \delta\,w^*_{k(i)} + \varepsilon_i$ is Gaussian in $\delta$, the full
    conditional is $\Normal(m_\delta P_\delta^{-1},\, P_\delta^{-1})$
    with precision $P_\delta = 1/100 + \sum_i (w^*_{k(i)})^2$
    and $m_\delta = \sum_i w^*_{k(i)} r_i$, under the prior
    $\delta \sim \Normal(0,\, 100)$.

  \item \textbf{LGCP intercept $\alpha_0$}:
    Random-walk MH: $\alpha_0' = \alpha_0 + \xi_{\alpha_0}$,
    $\xi_{\alpha_0} \sim \Normal(0,\, 0.2^2)$.
    Acceptance uses the LGCP log-likelihood
    \eqref{eq:lgcp_lik}--\eqref{eq:grid_approx} and prior
    $\alpha_0 \sim \Normal(0,\, 100)$.

  \item \textbf{LGCP covariate coefficients $\bgamma$}:
    Element-wise random-walk MH: $\gamma_j' = \gamma_j + \xi_j$,
    $\xi_j \sim \Normal(0,\, 0.15^2)$,
    updated one component at a time with acceptance based on the LGCP
    log-likelihood and prior $\gamma_j \sim \Normal(0,\, 100)$.

  \item \textbf{Store-level GP $\bm{w}^{\mathrm{obs}}$}:
    Sequential exact Gibbs, one observed location at a time. The probit
    likelihood contributes unit precision per site through the residual
    $z_i - \bx_i^\top\bbeta - \delta w^*_{k(i)}$, combined with the NNGP
    full-conditional formed from the parent and child nodes of the store-level
    neighbour graph.

  \item \textbf{Store-level GP variance $\sigma^2_{\mathrm{obs}}$}:
    Exact inverse-gamma Gibbs draw, of the same form as step~4 applied to
    $\bm{w}^{\mathrm{obs}}$ over the $n$ observed locations.

  \item \textbf{Store-level GP range $\phi_{\mathrm{obs}}$}:
    Single-parameter adaptive random-walk MH on $[0.3, 2.0]$, of the same form
    as step~5, with the acceptance ratio using the NNGP log-density of
    $\bm{w}^{\mathrm{obs}}$.

\end{enumerate}

\noindent
The other specifications are reductions of this cycle, obtained by deleting the
blocks they do not contain. The \textbf{PS Model (Model~3)} drops the
store-level field, i.e.\ steps~9--11 and the $w^{\mathrm{obs}}(\bs_i)$ terms in
steps~1--3 and~6. \textbf{Model~2} keeps the latent-variable, $\bbeta$ and
store-level-field updates (steps~1--2 and~9--11) but omits the shared field and
the location process, dropping steps~3--8. \textbf{Model~1} (non-spatial
probit) retains only steps~1--2. The \textbf{standalone LGCP} keeps only the
location-process updates---the grid-level field $\bm{w}^*$ with its
hyperparameters $(\sigma^2_*, \phi_*)$ and the LGCP parameters $(\alpha_0,
\bgamma)$, i.e.\ steps~3--5 and~7--8---with no latent variables, $\bbeta$,
$\delta$ or store-level field.

\medskip
\noindent\textbf{MCMC settings.}
For every model we run $60{,}000$ iterations, discard the first
$30{,}000$ as burn-in, and thin by $10$, retaining $3{,}000$ posterior
draws. The same configuration is used throughout to keep the models
directly comparable.
Inference is based on a single long chain, so the multi-chain Gelman--Rubin
$\widehat{R}$ statistic is unavailable. We report effective sample sizes (ESS)
and marginal posterior histograms as basic diagnostics
(Appendix~\ref{app:convergence}); these diagnostics are less informative than
a multi-chain assessment.
Posterior summaries (mean, standard deviation, 2.5\% and 97.5\%
quantiles) are reported for all parameters.

\medskip
\noindent\textbf{Identifiability.}
A sum-to-zero constraint is imposed on the grid GP after each sweep
($\bm{w}^* \leftarrow \bm{w}^* - \bar{w}^*$), with the absorbed mean
reallocated to the intercepts: $\delta\,\bar{w}^*$ is added to the probit
intercept $\beta_0$ and $\bar{w}^*$ to the LGCP intercept $\alpha_0$.
This removes the confounding between the GP mean and the intercepts---the
overall level $\bar{w}^*$ can otherwise be traded off against $\alpha_0$ in the
LGCP log-intensity (which carries $\bm{w}^*$ at unit loading) and against
$\beta_0$ in the probit---thereby preventing unbounded drift in the intercepts
and the GP mean and stabilizing the $\delta$ posterior \citep{Banerjee2015}.
(Because $\bm{w}^*$ enters the LGCP log-intensity with unit loading, the joint
model is not invariant under the sign flip
$(\bm{w}^*,\delta)\leftrightarrow(-\bm{w}^*,-\delta)$, so no such sign ambiguity
arises here.)

%% ============================================================
%% 4. STUDY AREA AND DATA
%% ============================================================
\section{Study Area and Data}
\label{sec:data}

\subsection{Mini-Supermarkets in Tokyo's 23 Special Wards}
\label{subsec:data_stores}

\begin{figure}[htbp]
  \centering
  \includegraphics[width=0.60\textwidth]{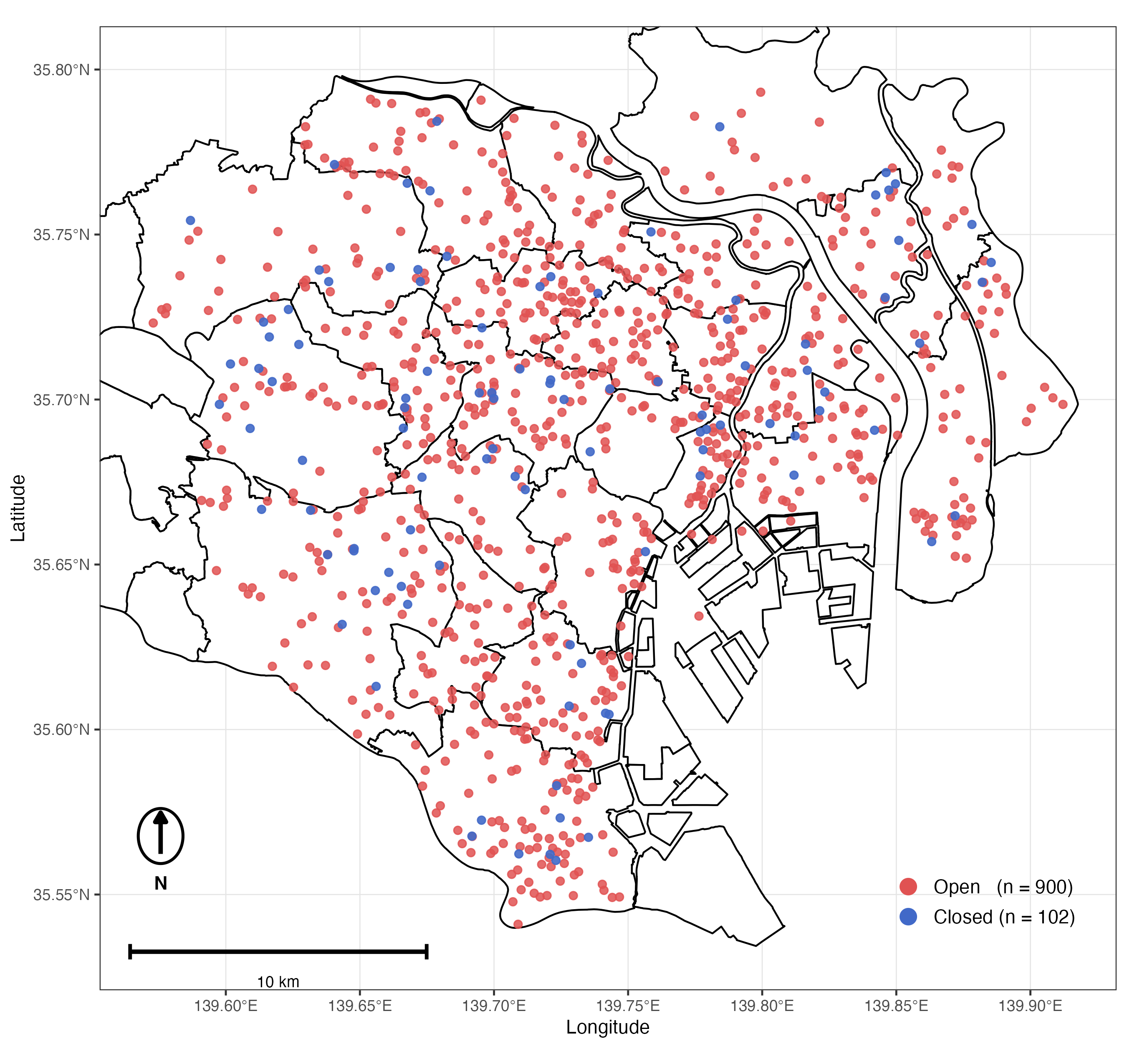}
  \caption{Mini-supermarket locations in Tokyo's 23 wards: operating (red) and closed (blue)}
  \label{fig:area}
\end{figure}

The dataset comprises $n = 1{,}002$ mini-supermarkets from three chains
active in Tokyo's 23 special wards as of October 31, 2025:
Maibasuketto, Maruetsu Petit, and Rico's.
We define a mini-supermarket as a small-format food retailer
($\leq 500\,\mathrm{m}^2$ floor area) primarily selling fresh produce
and daily necessities.

Outlet status was verified through a combination of official chain
websites, Google Maps, and street-view imagery.
A store is classified as \textit{closed} ($y_i = 0$; $n = 102$) if
it was permanently shuttered and not relocated or demolished for reasons
unrelated to business performance.
The remaining 900 stores are classified as \textit{operating}
($y_i = 1$), having been continuously open for at least one year.
The resulting closure rate is approximately 10.2\%, reflecting genuine
business attrition. Figure~\ref{fig:area} shows the mini-supermarket locations in Tokyo's 23 wards.
Closures due to relocation or redevelopment were identified via manual inspection of Google Maps Street View imagery and were excluded from the analysis.\footnote{Google Maps, Google LLC. https://maps.google.com (accessed October 31, 2025).}
Store coordinates (latitude and longitude) were obtained from chain
databases.

``LocationMind xPop'' Data refers to people flows data collected by individual location data sent from mobile phone under users' consent through provided by NTT DOCOMO, INC. The data is processed collectively and statistically in order to conceal the private information. Original location data is GPS data (latitude, longitude) sent at a frequency of every 5 minutes at the shortest interval and does not include information that specifies individuals.
\subsection{Explanatory Variables}
\label{subsec:covariates}

Table~\ref{tab:variables} lists the $p = 7$ explanatory variables
(excluding the intercept) used in the probit equation.
The three distance variables, the land price, and the daytime population
enter as $\log(x+1)$ transforms to compress their right-skewed
distributions; the
resulting variables are then standardized to zero mean and unit variance,
so that the coefficients are directly comparable in magnitude.
Standardization uses the means and standard deviations computed over the
integration grid, and the store-level covariates are rescaled with these
same grid-based parameters, so that the probit and LGCP covariate
equations share a common scale.

\begin{table}[htbp]
  \centering
  \caption{Explanatory variables used in the analysis. Covariates are
           standardized using the means and standard deviations computed over
           the integration grid.}
  \label{tab:variables}
  \small
  \begin{tabular}{lll}
    \toprule
     Variable & Definition & Unit \\
    \midrule
    
      Nearest\_Supermarket\_Distance\_m
      & Distance to nearest full supermarket & m \\
      Nearest\_ConvenienceStore\_Distance\_m
      & Distance to nearest convenience store & m \\
      Nearest\_Station\_Distance\_m
        & Distance to nearest rail station & m \\
      Stay\_Home
        & Share of dwell time attributed to residential use & \% \\
      Nearest\_Land\_Price
        & Nearest published land price & JPY/m$^2$ \\
      Resident\_Age\_Average\_4th\_New
        & Mean resident age (500-m 4th-level mesh) & years \\
      Daytime\_Pop
        & Daytime stay count & persons \\
    \bottomrule
  \end{tabular}
\end{table}

The mobility variable Stay\_Home is derived from the
LocationMind xPop dataset, which records the estimated proportions of time that
mobile devices registered in each mesh cell are attributed to
residential, commuting, and other activities.
Population data are drawn from the 2020 Japanese Population
Census.
Land prices are from the 2024 Ministry of Land, Infrastructure, Transport
and Tourism standard land price survey.

\subsection{Grid Data for Point Process Integration}
\label{subsec:grid}

\begin{figure}[htbp]
  \centering
  \includegraphics[width=0.60\textwidth]{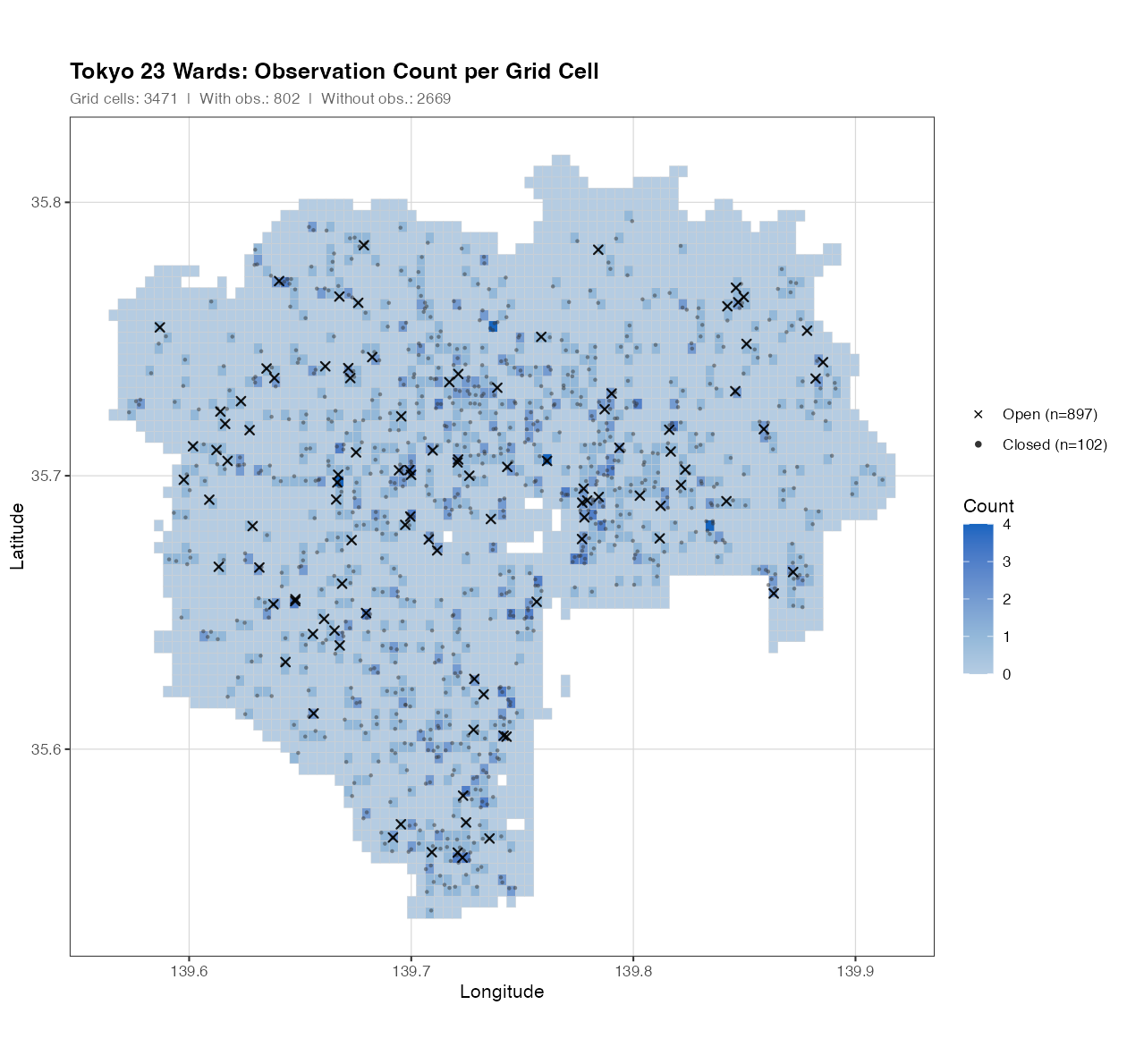}
  \caption{Observation count per grid cell in Tokyo's 23 wards}
  \label{fig:count}
\end{figure}

The LGCP likelihood integral \eqref{eq:lgcp_lik}--\eqref{eq:grid_approx}
requires a spatial discretization of the study domain.
We use a regular latitude--longitude grid of $K = 3{,}471$ points covering
Tokyo's 23 special wards, with a spacing of approximately $0.004^\circ$ in
each direction (about $0.4\,\text{km}$). Grid cells over the airport and the
reclaimed Tokyo Bay waterfront are removed so that they do not contribute
spurious intensity: the Haneda Airport area (latitude $35.540$--$35.570$,
longitude $139.755$--$139.900$) and three bay blocks (latitude
$35.570$--$35.610$, longitude $139.755$--$139.795$; latitude
$35.595$--$35.650$, longitude $139.770$--$139.820$; latitude
$35.620$--$35.665$, longitude $139.815$--$139.860$) are excluded, and the
same exclusions are applied to the observation set.
After these exclusions, the observation set entering the point-process and
survival likelihoods comprises $999$ stores ($897$ operating and $102$
closed): three operating stores lying on excluded cells are dropped from the
full sample of $1{,}002$.
Grid covariate values are extracted using the same data sources as the
store locations. Figure~\ref{fig:count} shows the counts and observed locations on the grids.
Consistent with the normalized domain of Section~\ref{subsec:ps_model}
($|\mathcal{D}| = 1$), each cell has area $A_k = 1/K$.
Each store is assigned to its nearest grid point for the LGCP likelihood
calculation; the number of stores per grid cell (count\_g) ranges from
0 to 4.

\section{Simulation Study}
\label{sec:simulation}

The real-data analysis returns $\hat\delta \approx 0$. Because a null result
is only informative if the estimator is capable of detecting preferential
sampling when it is present, we conduct a simulation study that (i) examines
parameter recovery under selected known-truth scenarios, (ii) illustrates how
sample size and class balance affect the precision of $\delta$, and
(iii) evaluates out-of-sample prediction of held-out closures.

\subsection{Data Generating Processes}
\label{sec:dgp}

Data are simulated from the Model~4 specification \eqref{eq:model4}, i.e.\ the
full joint model with both a shared grid GP $w^*$ and a store-level GP
$w^{\mathrm{obs}}$. Store locations are drawn from the LGCP intensity
\eqref{eq:lgcp} on the Tokyo grid, and binary survival outcomes are generated
from the probit model \eqref{eq:model4}. The data-generating parameters are
held fixed across scenarios: in the survival equation only the
nearest-supermarket distance loads on the outcome
($\beta_{\text{SM}} = 0.3$, all other $\beta_j = 0$); the LGCP coefficients
are $\bgamma = (0,\,-0.2,\,-0.3,\,0.2,\,-0.2,\,0.7,\,0)$ for (supermarket
distance, convenience-store distance, station distance, stay-home, resident
age, daytime population, land price); and the GP hyperparameters are
$\sigma^2_* = \sigma^2_{obs} = 0.2$, $\phi_* = \phi_{obs} = 1$.

We cross three factors in a $2\times 2\times 2$ design, yielding eight
scenarios:
\begin{itemize}[leftmargin=2em]
  \item sample size $n \in \{1000,\, 2000\}$;
  \item closure rate (prevalence of $y = 0$) $\in \{10\%,\, 30\%\}$, set
        through the survival intercept;
  \item shared-process loading $\delta \in \{0,\, 1\}$, contrasting a
        \textit{no-PS} regime ($\delta = 0$) against a \textit{strong-PS}
        regime ($\delta = 1$).
\end{itemize}
The $10\%$ closure rate mirrors the real data; the $30\%$ rate probes whether
the weak identification of $\delta$ in the application is driven by class
imbalance. Each scenario is fitted with the same Metropolis-within-Gibbs
sampler used for the real data.

\subsection{Parameter Recovery}
\label{sec:recovery}

Table~\ref{tab:sim_delta} reports the loading $\delta$, the parameter of
primary interest, across the eight reported scenarios. The 95\% credible
interval contains the true value in every reported run. When $\delta = 1$
(strong PS), the interval excludes zero in all four runs; when $\delta = 0$
(no PS), it includes zero in all four runs. The survival coefficient
$\beta_{\text{SM}}$ also has posterior means within roughly $0.05$ of the
truth in these runs. These results provide scenario-specific implementation
checks, but they are not a repeated-simulation assessment of frequentist
coverage or false-positive rates.

\begin{table}[htbp]
  \centering
  \caption{Recovery of the shared-process loading $\delta$ across the eight
           simulation scenarios (Model~4 data-generating process).
           ``PS detected'' indicates whether the 95\% credible interval
           excludes zero. All reported intervals contain the true value.}
  \label{tab:sim_delta}
  \small
  \begin{tabular}{ccccccc}
    \toprule
    $n$ & Closure & True $\delta$ & Post.\ mean & 95\% CrI & Covered & PS detected \\
    \midrule
    1000 & 10\% & 0 & \phantom{-}0.17 & $[-0.68,\ 0.88]$ & Yes & No  \\
    1000 & 10\% & 1 & \phantom{-}1.14 & $[\phantom{-}0.40,\ 2.09]$ & Yes & Yes \\
    1000 & 30\% & 0 & \phantom{-}0.41 & $[-0.03,\ 0.97]$ & Yes & No  \\
    1000 & 30\% & 1 & \phantom{-}1.18 & $[\phantom{-}0.52,\ 1.95]$ & Yes & Yes \\
    2000 & 10\% & 0 & \phantom{-}0.29 & $[-0.07,\ 0.71]$ & Yes & No  \\
    2000 & 10\% & 1 & \phantom{-}0.87 & $[\phantom{-}0.44,\ 1.32]$ & Yes & Yes \\
    2000 & 30\% & 0 & $-0.20$ & $[-0.55,\ 0.11]$ & Yes & No  \\
    2000 & 30\% & 1 & \phantom{-}0.96 & $[\phantom{-}0.59,\ 1.36]$ & Yes & Yes \\
    \bottomrule
  \end{tabular}
\end{table}

The width of the $\delta$ interval generally narrows as either the sample
size or the closure rate increases, consistent with the effective number of
closures being important for the precision of the shared-process loading. At
the real-data operating point ($n \approx 1000$, $10\%$ closure), the no-PS
interval is wide and centered near zero, illustrating the limited precision
available in a dataset with relatively few closures.

\subsection{In-Sample Discrimination of Closure Risk}
\label{sec:sim-prediction}

We additionally evaluate the in-sample discrimination of the closure
probabilities fitted by Model~4. The fitted probabilities are scored against
the simulated binary outcomes from the same sample. This exercise does not
directly measure recovery of the latent closure-probability surface; rather, it
measures how well fitted risks rank the realized closures in sample.
Table~\ref{tab:sim_auprc} reports the area under the precision--recall curve
(AUC-PR), together with the prevalence baseline. AUC-PR exceeds the baseline in
every reported scenario. Within each $(n, \text{closure})$ cell, the strong-PS
scenario ($\delta = 1$) also has a higher AUC-PR than the corresponding no-PS
scenario ($\delta = 0$), indicating greater in-sample discrimination when the
shared process contributes outcome-relevant signal.

\begin{table}[htbp]
  \centering
  \caption{In-sample discrimination of closure risk by Model~4, measured by
           AUC-PR against the simulated binary outcomes.
           ``Baseline'' is the closure prevalence.
           Within each $(n,\text{closure})$ cell, the strong-PS scenario
           ($\delta = 1$) outperforms the no-PS scenario ($\delta = 0$).}
  \label{tab:sim_auprc}
  \small
  \begin{tabular}{ccccc}
    \toprule
    $n$ & Closure & True $\delta$ & AUC-PR & Baseline \\
    \midrule
    1000 & 10\% & 0 & 0.630 & 0.097 \\
    1000 & 10\% & 1 & 0.788 & 0.099 \\
    1000 & 30\% & 0 & 0.750 & 0.290 \\
    1000 & 30\% & 1 & 0.897 & 0.309 \\
    2000 & 10\% & 0 & 0.478 & 0.111 \\
    2000 & 10\% & 1 & 0.552 & 0.096 \\
    2000 & 30\% & 0 & 0.747 & 0.294 \\
    2000 & 30\% & 1 & 0.846 & 0.301 \\
    \bottomrule
  \end{tabular}
\end{table}

\subsection{Out-of-Sample Predictive Performance}
\label{sec:sim-oos}

The in-sample exercise above scores fitted probabilities against outcomes from
the same simulated data. Because the simulation lets us control the number of
events by design, we also assess \emph{out-of-sample} prediction. We do not
perform the corresponding exercise for the smaller real-data sample
(Section~\ref{sec:results}). We focus
on the most data-rich design ($n = 2000$, $30\%$ closure), which leaves enough
held-out closures to estimate predictive performance reliably, and compare
Model~1 (non-spatial probit), Model~2 (store-level GP), and Model~4 (the full
joint specification with both the shared field $\delta w^{*}$ and a store-level
field $w^{\mathrm{obs}}$).

Each model is estimated on a training subsample from which the survival
\emph{labels} of the held-out stores are withheld; the store \emph{locations}
are never removed, so the LGCP location likelihood always sees the complete
spatial pattern and only the binary survival outcome is predicted. For each
held-out store $i$ we form a posterior-averaged plug-in closure probability
\begin{equation*}
  \hat p_i = \frac{1}{S}\sum_{s=1}^{S}
    \Big[1 - \Phi\!\big(\bx_i^{\top}\bbeta^{(s)}
      + \delta^{(s)}\,w^{*(s)}(\bg_{k(i)})
      + \widehat w^{\mathrm{obs},(s)}_{i}\big)\Big],
\end{equation*}
averaging over the $S$ retained posterior draws; since $y_i = 1$ denotes an operating store, closure corresponds to $z_i \le 0$ and hence to the complementary probability $1 - \Phi(\cdot)$. The shared grid field $w^{*}$
is read directly from the grid cell $\bg_{k(i)}$ containing the held-out store,
since that field already spans the whole domain; the store-level field is represented by its Gaussian-process conditional mean
$\widehat w^{\mathrm{obs},(s)}_{i}$ (kriging) from the training stores, with
the covariance parameters $(\sigma^{2}_{\mathrm{obs}},
\phi_{\mathrm{obs}})$ fixed at their posterior means. Thus the predictor uses
a kriging-mean plug-in rather than integrating over the conditional variance of
the held-out store-level field. Model~1 uses the fixed effect $\bx_i^{\top}\bbeta$ alone and
Model~2 omits the shared component $\delta w^{*}$. Treating closure ($y = 0$) as
the positive class, predictions are scored against the held-out labels by the
area under the precision--recall curve (AUC-PR) and the Brier score
\citep{Brier1950}, with the
held-out closure prevalence as the AUC-PR baseline. We consider two
complementary held-out designs: a \emph{random} hold-out of $200$ stores
($\sim\!10\%$ of $n$), which probes interpolation within the observed footprint,
and a contiguous \emph{spatial block}---a square region grown around the centre
of the study area until it contains roughly $60$ closed stores (here $232$
stores, $\sim\!12\%$ of $n$)---which probes extrapolation beyond the training
footprint and is the more demanding test of a spatial model.

\begin{figure}[htbp]
  \centering
  \includegraphics[width=\textwidth]{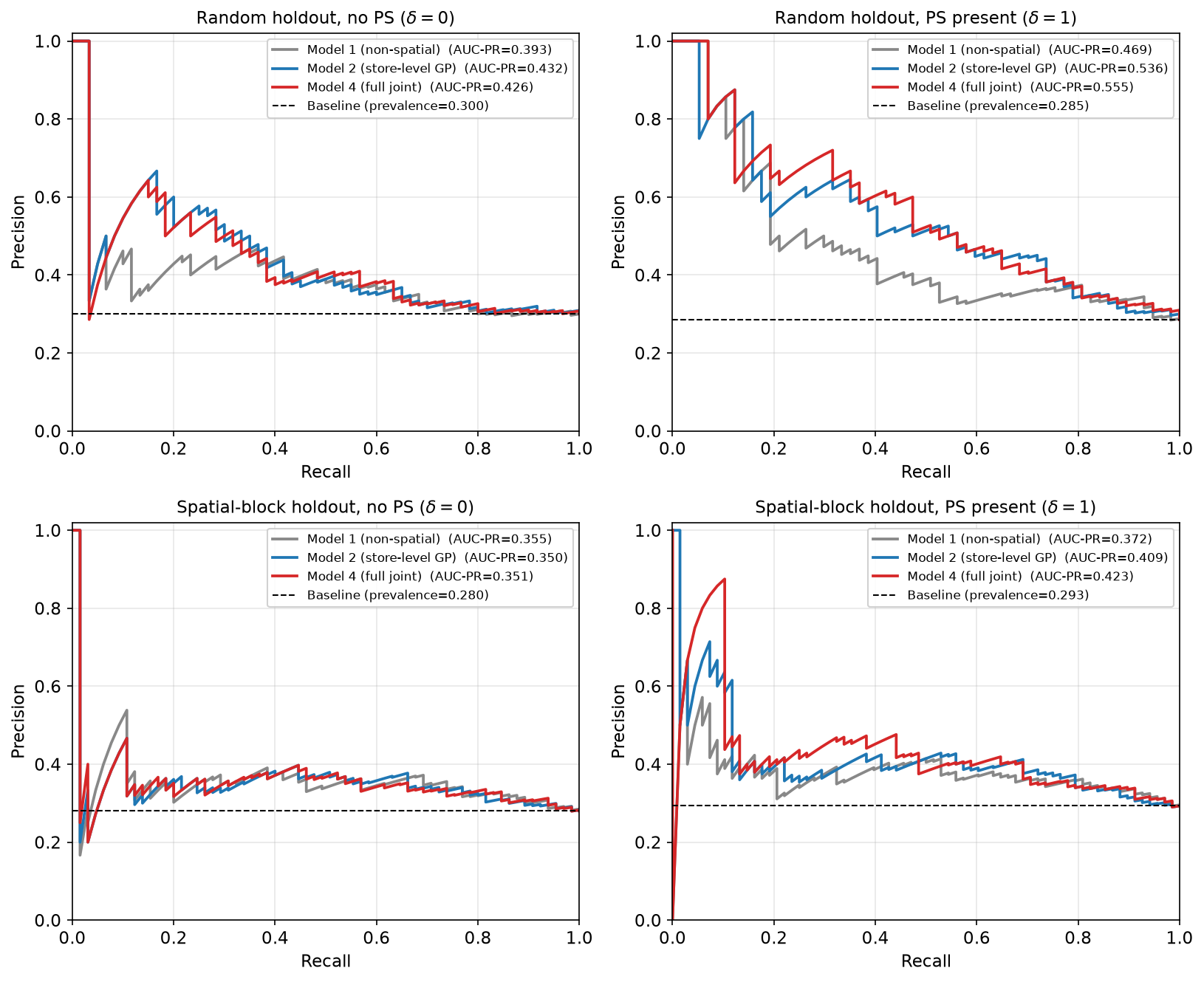}
  \caption{Out-of-sample precision--recall curves for the held-out closures
           ($n = 2000$, $30\%$ closure design). Rows: random vs.\ spatial-block
           held-out design; columns: no preferential sampling ($\delta = 0$)
           vs.\ preferential sampling present ($\delta = 1$). Each panel overlays
           Model~1 (non-spatial), Model~2 (store-level GP), and Model~4 (full
           joint); the dashed line is the held-out closure prevalence. The full
           joint Model~4 separates from the others only when preferential
           sampling is present.}
  \label{fig:sim_oos_pr}
\end{figure}

Table~\ref{tab:sim_oos} reports the results. Three findings stand out. First,
every model exceeds its prevalence baseline on the held-out sites in these
runs, indicating out-of-sample discrimination above the baseline. Second, when
preferential sampling is present ($\delta = 1$) the full joint Model~4 attains
the best AUC-PR and the lowest Brier score under both held-out designs (random:
$0.555$ vs.\ $0.536$ and $0.469$; block: $0.423$ vs.\ $0.409$ and $0.372$):
sharing the spatial field across the location and survival processes translates
into a real predictive gain precisely when that sharing exists in the truth.
Third, when there is no preferential sampling ($\delta = 0$) the three models are
essentially indistinguishable, and the extra shared field neither helps nor
hurts---the simulated analogue of the real-data result that the joint structure
adds little once $\delta \approx 0$. As expected, the spatial-block design
(extrapolation) is uniformly harder than the random design (interpolation),
compressing all models toward the baseline.
Even in this harder design, the full joint Model~4 retains an edge in the
reported $\delta=1$ run
under preferential sampling, for an instructive reason: because store
\emph{locations} are never withheld, the shared grid field $w^{*}$ stays
well-identified by the point-process likelihood even inside the held-out
block, whereas the store-level field $w^{\mathrm{obs}}$ must be extrapolated
from distant training stores and contributes little there. When
$\delta \neq 0$, the location-informed $w^{*}$ therefore carries
survival-relevant spatial structure into the extrapolation region that a
purely survival-driven field cannot recover---which is precisely why Model~4
outperforms the others in the spatial-block, $\delta = 1$ cell.
Figure~\ref{fig:sim_oos_pr} shows the corresponding precision--recall curves:
the separation of Model~4 (red) from the others is visible only in the
$\delta = 1$ panels, while under $\delta = 0$ the three curves overlap.

\begin{table}[htbp]
  \centering
  \caption{Out-of-sample predictive performance in the $n = 2000$, $30\%$
           closure design, by held-out design. Models are estimated on the
           training subsample and scored on held-out sites by AUC-PR (higher is
           better) and the Brier score (lower is better); ``Baseline'' is the
           held-out closure prevalence. Under preferential sampling
           ($\delta = 1$) the full joint Model~4 has the best AUC-PR in both
           designs and the best (or tied-best) Brier score; with no preferential
           sampling ($\delta = 0$) the models are essentially tied.}
  \label{tab:sim_oos}
  \small
  \begin{tabular}{llccccc}
    \toprule
    Held-out & True $\delta$ & Model & AUC-PR & Baseline & Brier \\
    \midrule
    \multirow{6}{*}{Random}
      & 0 & 1 (non-spatial)      & 0.393 & 0.300 & 0.209 \\
      & 0 & 2 (store-level GP)   & 0.432 & 0.300 & 0.206 \\
      & 0 & 4 (full joint)       & 0.426 & 0.300 & 0.207 \\
      & 1 & 1 (non-spatial)      & 0.469 & 0.285 & 0.188 \\
      & 1 & 2 (store-level GP)   & 0.536 & 0.285 & 0.177 \\
      & 1 & 4 (full joint)       & \textbf{0.555} & 0.285 & \textbf{0.174} \\
    \midrule
    \multirow{6}{*}{Spatial block}
      & 0 & 1 (non-spatial)      & 0.355 & 0.280 & 0.205 \\
      & 0 & 2 (store-level GP)   & 0.350 & 0.280 & 0.206 \\
      & 0 & 4 (full joint)       & 0.351 & 0.280 & 0.205 \\
      & 1 & 1 (non-spatial)      & 0.372 & 0.293 & 0.206 \\
      & 1 & 2 (store-level GP)   & 0.409 & 0.293 & 0.204 \\
      & 1 & 4 (full joint)       & \textbf{0.423} & 0.293 & \textbf{0.204} \\
    \bottomrule
  \end{tabular}
\end{table}

\section{Results}
\label{sec:results}

This section reports the results of applying the models to the \emph{real data}---the
mini-supermarkets observed across Tokyo's 23 special wards described in
Section~\ref{sec:data}. All estimates, credible intervals, and model-fit
diagnostics below are obtained from this single observed dataset and are evaluated
\emph{in sample}: we assess how well each specification fits the data it was
estimated on, rather than its ability to predict held-out observations. We restrict
the real-data analysis to in-sample model fit because the dataset is limited in
size---and, in particular, contains only $n_{\text{closed}} = 102$ closures against
$897$ surviving stores in the analysis sample. A training/test split would leave too few closures in the
held-out set to estimate held-out performance reliably and would further weaken the
already limited power to identify the loading $\delta$. Scenario-based checks under a known data-generating mechanism are therefore
reported in the simulation study in
Section~\ref{sec:simulation}, which examines recovery of $\delta$ in the
reported scenarios, evaluates in-sample discrimination, and predicts genuinely
held-out closures (Section~\ref{sec:sim-oos}).

\subsection{LGCP Coefficients}
\label{subsec:coef_lgcp}

\begin{figure}[htbp]
  \centering
  \includegraphics[width=0.7\textwidth]{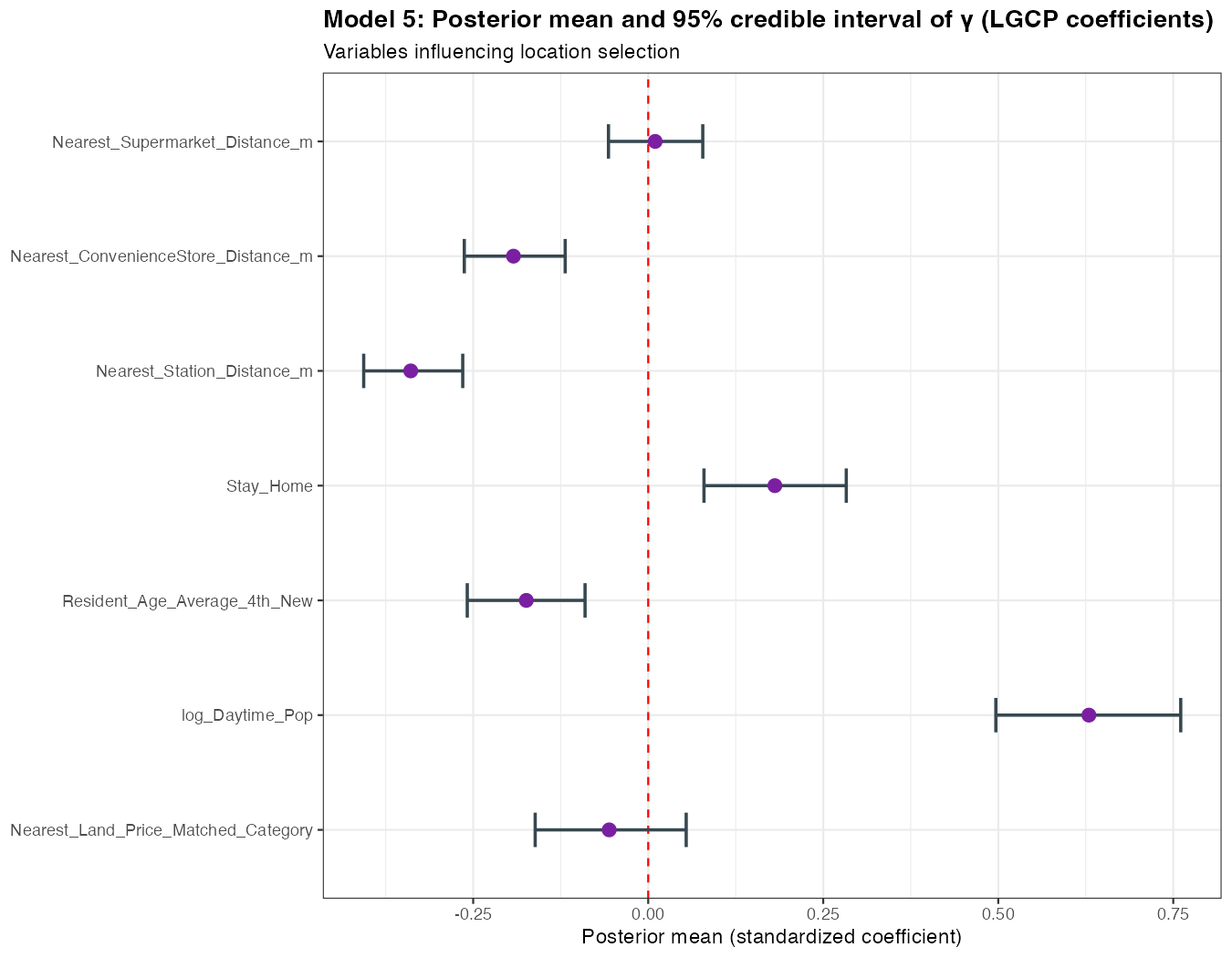}
  \caption{Forest plot of posterior means and 95\% credible intervals for the
           LGCP covariate coefficients $\bm{\gamma}$ (standalone LGCP model).
           All covariates are standardized.}
  \label{fig:posteriorlgcp}
\end{figure}

Figure~\ref{fig:posteriorlgcp} shows posterior means and $95\%$ credible intervals for $\bm{\gamma}$ in LGCP. The significant covariates (95\% credible intervals excluding zero) are the distance to the nearest convenience store, the distance to the nearest station, the stay-home share, the average resident age, and the daytime population. We discuss the interpretation for these estimates.

The strongly negative estimate for station distance indicates that stores are more likely to be opened in closer proximity to stations, suggesting a strong preference for station-adjacent locations. Additionally, the strongly positive estimate for the stay-home share suggests that mini-supermarkets are more likely to be established in areas with a high share of estimated residential dwell time, reflecting a clear orientation toward residential neighborhoods. Furthermore, the strongly positive estimate for daytime population indicates that a larger daytime population increases the likelihood of store entry; this variable has the largest positive posterior mean in the model. These results indicate a concentration of stores in residential areas near stations with high daytime populations (e.g., densely populated residential districts surrounding the city center). The fact that the daytime-population coefficient is the largest
($\gamma \approx +0.6$) is consistent with operators favoring areas with a
high volume of potential customers, including commuters and students, rather
than focusing solely on local residents.
Given that mini-supermarkets are typically concentrated in urban centers, these findings are highly consistent with established store development patterns in the industry. 

The negative estimate for average resident age indicates that stores are more likely to be opened in areas with a younger population (a trend favoring locations with fewer elderly residents). While this might suggest a preference for areas with better commuting environments, such as those near stations, the effect is weaker than that of the other variables, so this interpretation should be treated cautiously.

The distance to the nearest convenience store carries a significantly negative
coefficient ($\gamma \approx -0.19$), indicating that mini-supermarkets tend
to open close to convenience stores. This association may reflect both formats
being drawn to the same high-footfall, station-adjacent residential pockets,
rather than direct complementarity. By contrast, the distance to the nearest full supermarket is not significant, suggesting that the presence of larger competitors is not a primary factor in store-entry decisions, or that an offsetting effect is at play whereby areas with high competition also possess sufficient demand to compensate.

These location-process estimates are essentially unchanged whether the LGCP is
fitted on its own or jointly with the survival outcomes: the coefficients from
the PS Model and Model~4 differ from the standalone LGCP by at most
$\approx 0.02$ in magnitude, with no change in sign or interval-based
significance. In this application, the point-process likelihood therefore
dominates the identification of $w^{*}$ and $\bm{\gamma}$; because the
posterior for $\delta$ is centered near zero, the survival likelihood adds
little information about the location field. The standalone LGCP coefficients
are consequently representative of the joint-model estimates in this dataset.

\subsection{Probit Regression Coefficients}
\label{subsec:coef_results}

\begin{table}[htbp]
  \centering
  \caption{Posterior means (95\% credible intervals) for $\bm{\beta}$ of probit
           regression coefficients. SM Dist and CS Dist denote the distance to
           the nearest supermarket and convenience store, respectively.
           Bold entries indicate significant covariates
           (95\% credible interval excluding zero). All covariates are standardized.}
  \label{tab:beta_comparison}
  \small
  \begin{tabular}{lcccc}
    \toprule
    & Model 1 & Model 2 & Model 3 & Model 4 \\
    Variables & (Probit) & ($+w^{\mathrm{obs}}$) & ($+w^{*}$) & ($+w^{\mathrm{obs}}+w^{*}$) \\
    \midrule
    Intercept
      & \textbf{1.53 [1.32, 1.73]} & \textbf{1.74 [1.48, 2.06]} & \textbf{1.52 [1.31, 1.77]} & \textbf{1.78 [1.49, 2.12]}  \\
    SM Dist
      & \textbf{0.24 [0.14, 0.35]} & \textbf{0.27 [0.15, 0.39]} & \textbf{0.25 [0.14, 0.35]} & \textbf{0.27 [0.15, 0.39]} \\
    CS Dist
      & 0.03 [-0.09, 0.15] & 0.03 [-0.10, 0.17] & 0.03 [-0.09, 0.14] & 0.03 [-0.11, 0.16] \\
    Station Dist
      & 0.05 [-0.05, 0.16] & 0.06 [-0.07, 0.20] & 0.05 [-0.06, 0.17] & 0.07 [-0.06, 0.20] \\
    Stay Home
      & -0.04 [-0.21, 0.13] & -0.06 [-0.27, 0.14] & -0.05 [-0.24, 0.13] & -0.07 [-0.27, 0.14] \\
    Land Price
      & -0.03 [-0.19, 0.13] & -0.04 [-0.23, 0.15] & -0.04 [-0.20, 0.13] & -0.05 [-0.24, 0.16] \\
    Resident Age
      & 0.09 [-0.07, 0.26] & 0.10 [-0.10, 0.31] & 0.09 [-0.07, 0.27] & 0.10 [-0.10, 0.30] \\
    Daytime Pop
      & -0.02 [-0.25, 0.21] & -0.04 [-0.32, 0.24] & -0.05 [-0.30, 0.19] & -0.04 [-0.33, 0.24] \\
    $\delta$
      & --- & --- & 0.26 [-0.78, 1.05] & -0.12 [-0.95, 0.75] \\
    \midrule
    \multicolumn{5}{l}{\textit{Spatial hyperparameters}} \\
    $\sigma_{\mathrm{obs}}^2$
      & --- & 0.32 [0.15, 0.66] & --- & 0.33 [0.14, 0.67] \\
    $\phi_{\mathrm{obs}}$
      & --- & 1.26 [0.39, 1.96] & --- & 1.21 [0.37, 1.96] \\
    $\sigma_{*}^2$
      & --- & ---  & 0.16 [0.10, 0.25] & 0.17 [0.11, 0.26] \\
    $\phi_{*}$
      & --- & --- & 0.54 [0.30, 1.36] & 0.67 [0.32, 1.31] \\
    \bottomrule
  \end{tabular}
\end{table}

Table~\ref{tab:beta_comparison} compares posterior means and 95\%
credible intervals (CrI) for $\bbeta$ across the four models. The only consistently significant covariate is the distance to the nearest supermarket (SM Dist); the intercept is also strongly positive in every specification. One possible explanation is restriction of range: the stores that actually
opened are concentrated in areas favored by the covariates associated with the
LGCP intensity, leaving less between-store variation in those covariates than
exists across the full grid. Together with the limited number of closures and
correlation among covariates, this may contribute to their wide posterior
intervals in the survival equation. SM Dist, which is not credibly associated
with the LGCP entry intensity, is the only covariate whose interval excludes
zero consistently in the survival models. The strongly positive intercept reflects the high overall survival rate (approximately 90\%), so that established mini-supermarkets generally have a high baseline probability of survival.

The positive estimate for SM Dist indicates that the probability of survival increases as the distance to the nearest supermarket grows. This suggests that full-scale supermarkets may act as competitors to mini-supermarkets. None of the remaining covariates---including the distance to convenience stores, the stay-home share, the daytime population, the resident age, and the land price---reaches significance in the survival equation, indicating that the spatial structure governing store \textit{location} (captured by the LGCP) does not translate into a comparable structure in store \textit{survival}.

Models~3 and~4 include the loading parameter $\delta$, which couples the
shared grid GP to the survival equation. In both models the 95\% credible
interval includes zero: the PS Model gives $\hat\delta = 0.26$ (95\% CrI
$[-0.78, 1.05]$) and Model~4 gives $\hat\delta = -0.12$ (95\% CrI
$[-0.95, 0.75]$). We therefore find no clear evidence of preferential sampling
in this dataset. This should not be read as evidence that the loading is exactly
zero: the intervals remain wide, in part because the analysis contains only
$n_{\text{closed}} = 102$ closures. In the reported matched simulation
($n=1000$, 10\% closure), the interval excludes zero in the $\delta=1$ run
and includes zero in the $\delta=0$ run (Table~\ref{tab:sim_delta}), showing
that the implementation can distinguish these two particular scenarios. It
does not establish a general power curve or rule out weak-to-moderate loadings
in the real data. Importantly, the posterior estimates of $\bbeta$ are nearly
unchanged across Models~1--4 (Table~\ref{tab:beta_comparison}), so the
substantive coefficient conclusions are not sensitive to including the shared
field within the specifications considered here.

The joint model is useful even when the loading is not clearly separated from
zero: it directly assesses whether the residual field governing store locations
also contributes to survival. Here the interval-based assessment is
inconclusive about small or moderate loadings, but the stability of the
regression coefficients across the PS and non-PS models shows that the main
covariate conclusions are robust to this modeling extension.

\subsection{Model Fit}
\label{subsec:performance}
\begin{figure}[htbp]
  \centering
  \includegraphics[width=0.7\textwidth]{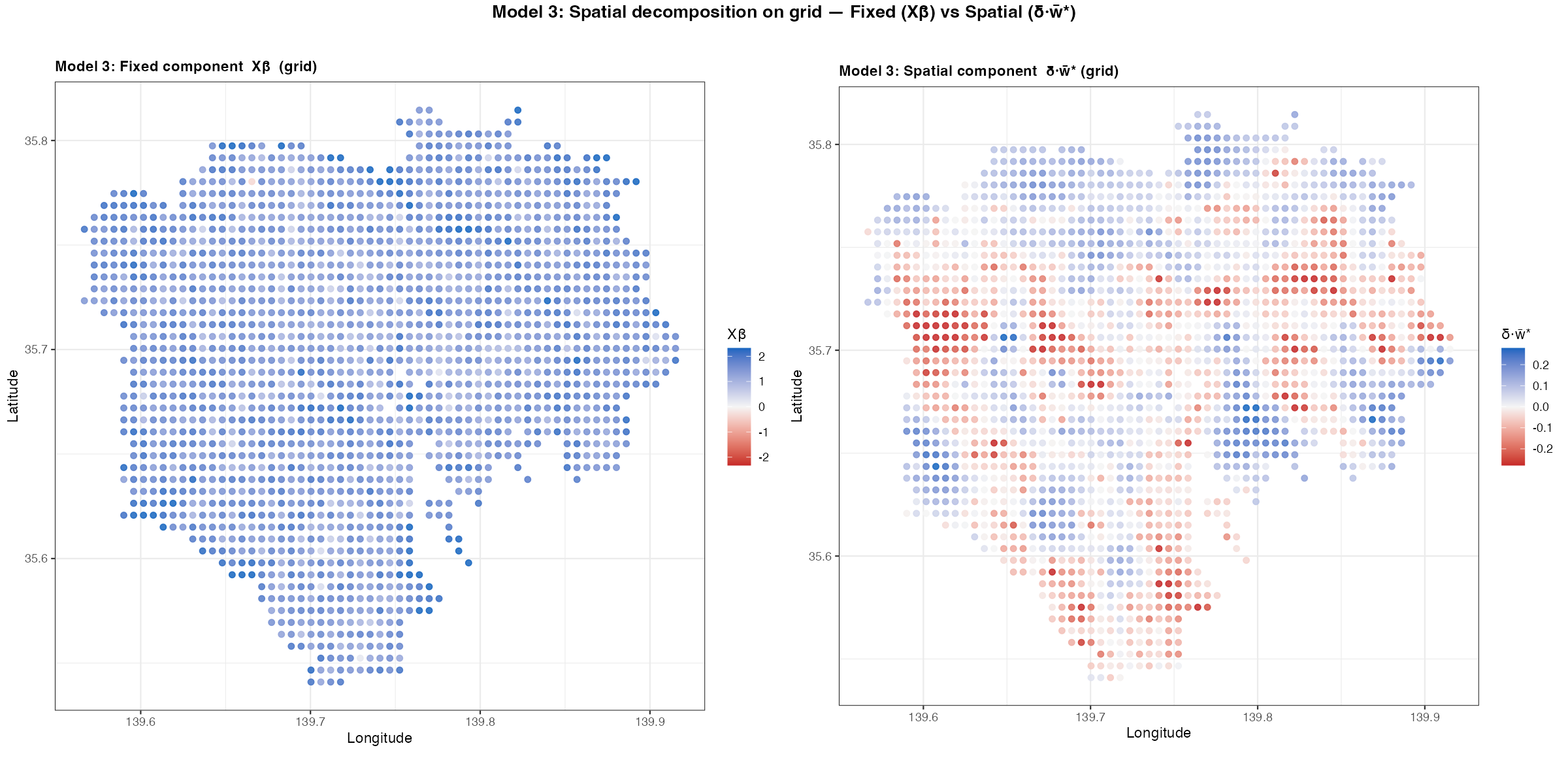}
  \caption{Posterior mean surface of $\bm{x}^{\top}\bm{\beta}$ (left) and $\delta w^{*}$ (right) in Model 3}
  \label{fig:decomp3}
\end{figure}

Figure~\ref{fig:decomp3} maps the posterior means of the covariate effect $\bm{x}^{\top}\bm{\beta}$ and the shared GP contribution $\delta w^*(\bg_k)$ across the 3,471 grid points covering Tokyo's 23 wards.
The covariate term dominates the survival surface, whereas the shared-field term $\delta w^{*}$ is small in magnitude: because $\hat\delta \approx 0.26$ with a credible interval spanning zero, the shared spatial field contributes little to the fitted closure probabilities and shows no coherent large-scale structure. This is the spatial counterpart of the $\delta \approx 0$ finding---the latent field that organizes store \textit{locations} leaves almost no imprint on store \textit{survival}.

\begin{figure}[htbp]
  \centering
  \includegraphics[width=1\textwidth]{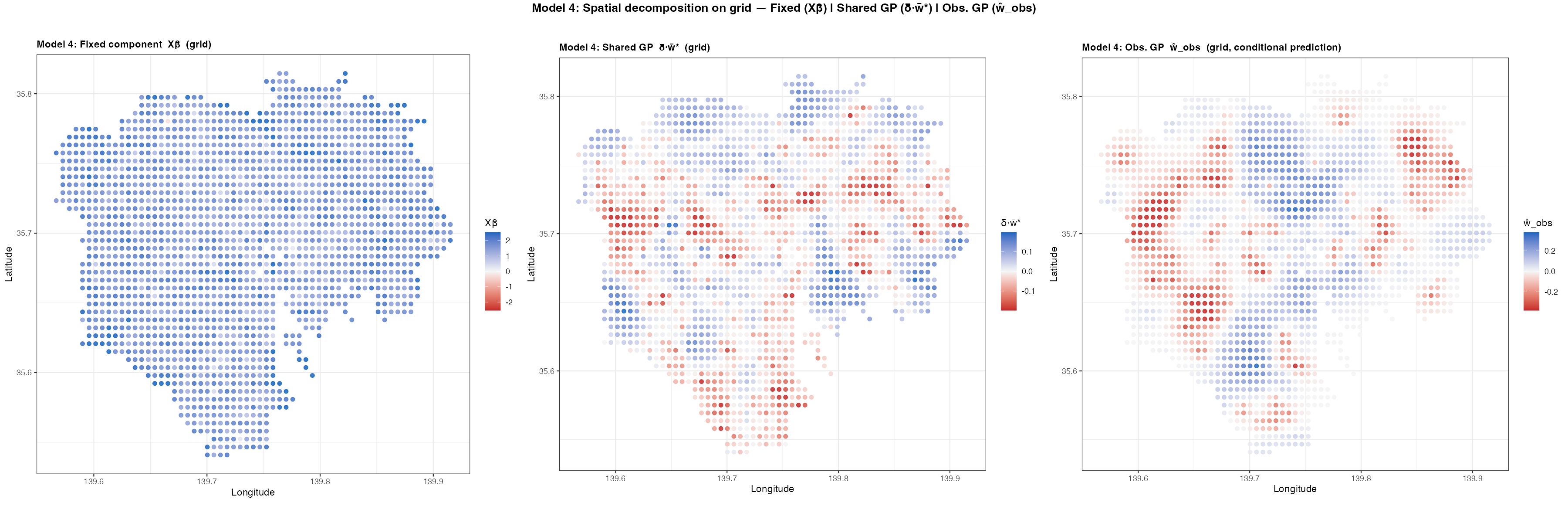}
  \caption{Posterior mean surface of $\bm{x}^{\top}\bm{\beta}$ (left), $\delta w^{*}$ (middle) and $w^{\mathrm{obs}}$ (right) in Model 4}
  \label{fig:decomp4}
\end{figure}

Figure~\ref{fig:decomp4} displays the posterior means of $\bm{x}^{\top}\bm{\beta}$, $w^{\mathrm{obs}}$, and $\delta w^{*}$ for Model 4. The covariate effects again dominate and the $\delta w^{*}$ term remains negligible ($\hat\delta \approx -0.12$, and of opposite sign to Model 3, underscoring that it is indistinguishable from zero), while the store-level field $w^{\mathrm{obs}}$ shows gradual spatial clustering that is independent of the location process. The fit improvement in Model 4 therefore comes from $w^{\mathrm{obs}}$ rather than from the shared, location-coupled field.

\begin{figure}[htbp]
  \centering
  \includegraphics[width=0.60\textwidth]{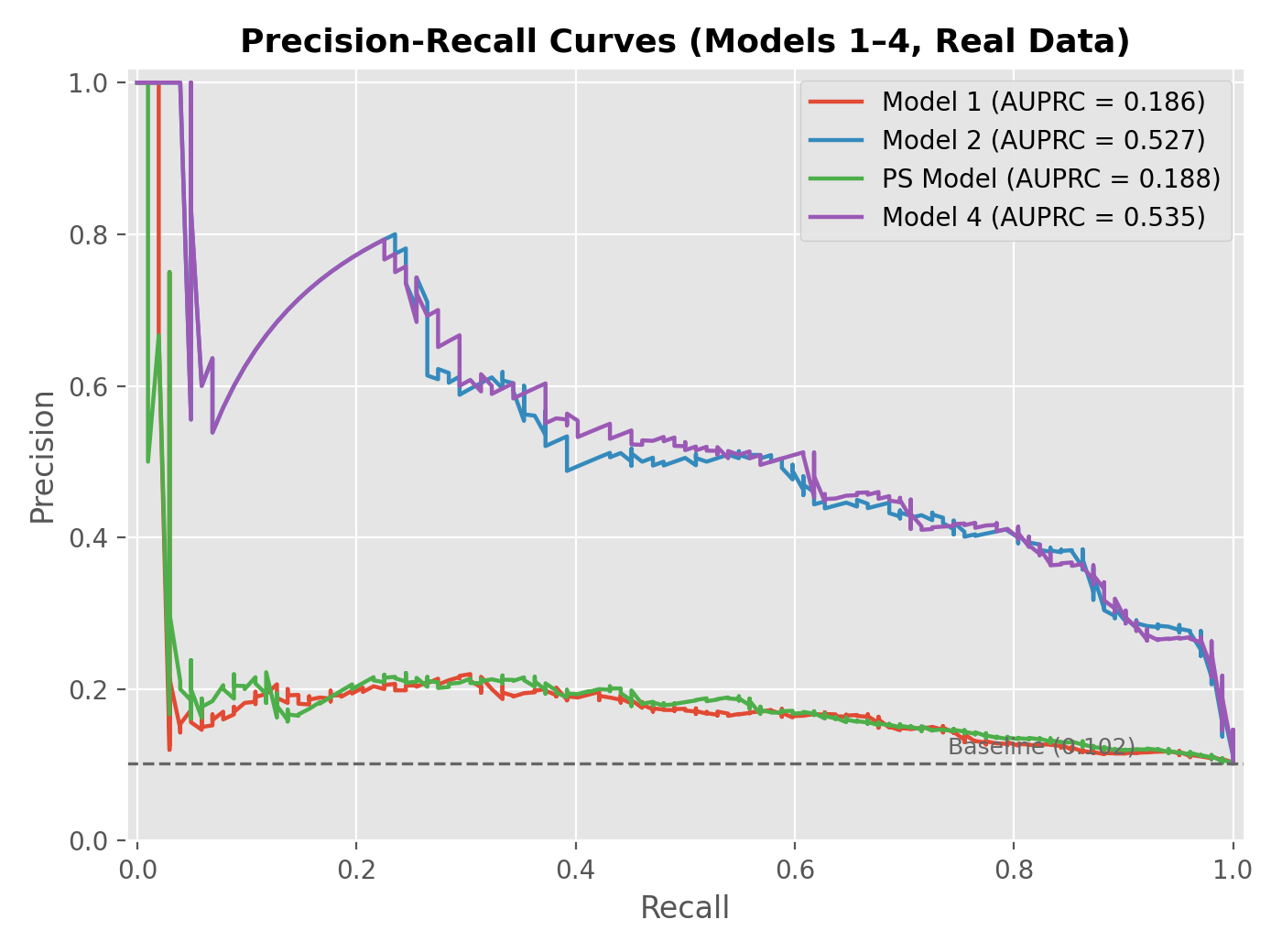}
  \caption{Precision--Recall curves for Models~1--4.
           Dotted horizontal line: random classifier baseline
           (prevalence $= 102/999 \approx 0.102$).
           In-sample AUC-PR values: Model~1 = 0.186; Model~2 = 0.527;
           PS Model = 0.188; Model~4 = 0.535.}
  \label{fig:pr}
\end{figure}

Figure~\ref{fig:pr} shows Precision--Recall (PR) curves for all four
models evaluated on the 999-store analysis sample using the posterior-mean
closure probabilities. Due to the limited number of observations for $y=0$, we did not perform a training/test split; instead, we focused on evaluating the in-sample fit.
We use the PR curve because it is generally more informative
than the ROC curve under substantial class imbalance \citep{Saito2015}; see
also \citet{Davis2006} for the formal relationship between ROC and PR spaces.

The relative ordering of the four curves reinforces the $\delta \approx 0$ conclusion. The PS Model (AUC-PR $= 0.188$) barely improves on the non-spatial Model~1 (AUC-PR $= 0.186$): adding the location-coupled channel $\delta w^{*}$ alone contributes essentially nothing to in-sample fit. By contrast, the two models that carry the store-level field $w^{\mathrm{obs}}$---Model~2 (AUC-PR $= 0.527$) and Model~4 (AUC-PR $= 0.535$)---show markedly better in-sample discrimination, and Model~4 adds little beyond Model~2 despite also including $\delta w^{*}$. The observed in-sample fit gain is therefore associated with $w^{\mathrm{obs}}$, a spatial effect that is estimated independently of the location process, rather than with the shared field that couples location and survival. Because the PR curves in Figure~\ref{fig:pr} and these AUC-PR values are in-sample summaries, the gain attributable to the flexible store-level field should be read as improved in-sample fit rather than as demonstrated out-of-sample predictive gain. Together these comparisons are consistent with the absence of a clearly detected shared location--survival field in the present data.

%% ============================================================
%% 6. CONCLUSION
%% ============================================================
\section{Conclusion}
\label{sec:conclusion}

Our results indicate that preferential sampling is \textit{not} detectable
in Tokyo's mini-supermarket data: the loading $\delta$ that couples the
shared spatial field to survival is statistically indistinguishable from
zero in both the PS Model and Model~4.
Although store locations are highly structured in space---as the LGCP
intensity makes clear---the residual location field does not have a clearly
identified association with survival in the present data.
The PS correction therefore leaves coefficient inference essentially
unchanged relative to the non-spatial baseline, and the gains in model fit
from the spatial specifications come from the store-level field $w^{\mathrm{obs}}$,
which is estimated independently of the location process, rather than from
the location-coupled shared field.
The reported simulation at the real-data operating point ($n \approx 1000$
and a 10\% closure rate) separates the selected $\delta=0$ and $\delta=1$
scenarios, but it is not a repeated-simulation power study. The real-data
credible intervals therefore leave weak-to-moderate loadings unresolved.
The most direct robustness result is that the posterior estimates of the
survival coefficients remain essentially unchanged across the PS and non-PS
specifications. Thus, within the models considered here, allowing for a shared
location field does not alter the substantive covariate conclusions.

The extension of the \citet{Diggle2010} PS framework to binary probit
responses, implemented via the \citet{AlbertChib1993} data augmentation
and the \citet{Datta2016} NNGP approximation, provides a computationally
tractable approach that avoids reliance on general-purpose probabilistic
programming software and scales to
the $n \approx 1{,}000$--$2{,}000$ range typical of urban retail
datasets.
The grid GP is updated by single-site random-walk Metropolis--Hastings,
with the Poisson (LGCP) and probit contributions evaluated exactly at each
grid point; this keeps every update local and inexpensive while avoiding any
linearization of the intensity surface.

In the survival equation, the single robust effect is the
distance to the nearest full supermarket: its positive probit coefficient
($\hat\beta_{\text{SM}} \approx 0.25$) implies that stores located closer to
full supermarkets face higher closure risk, consistent with the strongly
localized competitive effects of large-format grocery entry documented by
\citet{EllicksonGrieco2013}.
The remaining covariates that shape \textit{where} mini-supermarkets are
sited---proximity to stations and convenience stores, residential dwell
time, daytime population, and resident age, all significant in the LGCP
intensity---do not carry significant survival effects once location is
accounted for, reinforcing the interpretation that the spatial logic of
entry differs from the spatial logic of survival.
The prominence of station proximity and daytime population in
the LGCP is broadly consistent with previous evidence that geographic context
and urban mobility patterns are informative for retail-location performance
\citep{Karamshuk2013}.

Several limitations warrant attention.
First, closure status is determined by a cross-sectional snapshot. The model
does not incorporate opening dates, time at risk, or closure times, and the
current covariates may not coincide with the conditions prevailing when older
stores were opened. A time-to-event model with historically aligned covariates
would address these limitations.
Second, the closure classification relies partly on manual verification of
street-view imagery, introducing potential measurement error.
Third, the 10.2\% closure rate creates class imbalance and limits the precision
with which the PS loading $\delta$ can be estimated; for in-sample
discrimination we therefore report PR curves rather than ROC curves.
Fourth, inference is based on a single MCMC chain. ESS values and marginal
histograms provide only limited convergence diagnostics, and several regression
coefficients have relatively low ESS.
Fifth, while the NNGP approximation greatly reduces computational cost, it
introduces approximation error in addition to Monte Carlo error; we use
$m = 15$ neighbors, a commonly adopted choice that has yielded accurate
approximations in previous NNGP studies \citep{Datta2016,Finley2019}.
Finally, the analysis is confined to three retail chains; generalizing to
independent retailers or other Asian cities would broaden the external
validity of the findings.

This paper develops and applies a Bayesian hierarchical model for binary
retail survival outcomes that jointly accounts for (i) preferential
sampling in store location decisions, (ii) spatial autocorrelation in
latent survival potential, and (iii) the computational constraints of
urban-scale datasets.
The PS Model integrates a log-Gaussian Cox process for store location
with a probit regression for survival, sharing a Nearest-Neighbor
Gaussian Process spatial random effect, and is estimated via a
Metropolis-within-Gibbs sampler without relying on general-purpose
probabilistic programming software.

Applied to the 999-store analysis sample in Tokyo's 23 special wards, the
model finds no detectable preferential sampling ($\hat{\delta} \approx 0$,
with a 95\% credible interval spanning zero). Although this does not rule out
weak-to-moderate preferential sampling, the survival-coefficient estimates are
essentially unchanged across the PS and non-PS specifications.
Distance from the nearest full-scale supermarket is the single covariate whose
credible interval consistently excludes zero in the survival models. The other
spatial covariates are associated with the LGCP entry intensity but are not
clearly associated with survival in the fitted probit models.

Future work could extend the framework in two directions.
First, this analysis uses a binary operating-versus-closed indicator. If
opening and closure times could be measured accurately and more closures were
observed, a spatial time-to-event model would be a promising alternative.
Second, applying the model to multi-city data---including Seoul, Singapore,
and Osaka---would test the generalizability of the Tokyo findings and
contribute to the comparative urban retail geography literature.

%% ============================================================
%% CREDIT AUTHOR STATEMENT
%% ============================================================
\section*{CRediT authorship contribution statement}

\textbf{Akitoshi Kanetaka:} Conceptualization, Data curation, Methodology,
Software, Formal analysis, Investigation, Visualization, Writing -- original
draft. \textbf{Shinichiro Shirota:} Supervision, Methodology, Formal analysis
(preferential sampling model), Funding acquisition, Writing -- review \& editing.

%% ============================================================
%% DECLARATION OF COMPETING INTEREST
%% ============================================================
\section*{Declaration of competing interest}

The authors declare that they have no known competing financial interests or
personal relationships that could have appeared to influence the work reported
in this paper.

%% ============================================================
%% ACKNOWLEDGMENTS
%% ============================================================
\section*{Acknowledgments}

This research was supported in part by data provided by LocationMind Inc.
This work was supported by JSPS KAKENHI Grant Number 23K16848 and 26K14728.

%% ============================================================
%% DATA AVAILABILITY STATEMENT
%% ============================================================
\section*{Data Availability Statement}

Store location and status data are available from the corresponding author
upon reasonable request.
Mobility data (LocationMind xPop) are subject to a data-sharing agreement
with LocationMind Inc.\ and cannot be publicly released.
Socioeconomic and land price data are publicly available from the
Statistics Bureau of Japan and the Ministry of Land, Infrastructure,
Transport and Tourism.

%% ============================================================
%% DECLARATION OF GENERATIVE AI
%% ============================================================
\section*{Declaration of generative AI and AI-assisted technologies in the manuscript preparation process}

During the preparation of this work, the authors used ChatGPT (OpenAI) and
Claude (Anthropic) to assist with language editing and improving readability.
After using these tools, the authors reviewed and edited the content as needed
and take full responsibility for the content of the publication.

%% ============================================================
%% REFERENCES
%% ============================================================
\bibliographystyle{elsarticle-harv}
\bibliography{references}

@article{Diggle2010,
  author    = {Diggle, Peter J. and Menezes, Raquel and Su, Ting-li},
  title     = {Geostatistical inference under preferential sampling},
  journal   = {Journal of the Royal Statistical Society: Series C
               (Applied Statistics)},
  year      = {2010},
  volume    = {59},
  number    = {2},
  pages     = {191--232},
  doi       = {10.1111/j.1467-9876.2009.00701.x}
}

@article{Pati2011,
  author    = {Pati, Debdeep and Reich, Brian J. and Dunson, David B.},
  title     = {Bayesian geostatistical modelling with informative
               sampling locations},
  journal   = {Biometrika},
  year      = {2011},
  volume    = {98},
  number    = {1},
  pages     = {35--48},
  doi       = {10.1093/biomet/asq067}
}

@article{Gelfand2012,
  author    = {Gelfand, Alan E. and Sahu, Sujit K. and Holland, David M.},
  title     = {On the effect of preferential sampling in spatial prediction},
  journal   = {Environmetrics},
  year      = {2012},
  volume    = {23},
  number    = {7},
  pages     = {565--578},
  doi       = {10.1002/env.2169}
}

@article{Cecconi2016,
  author    = {Cecconi, Lorenzo and Grisotto, Laura and Catelan, Dolores
               and Lagazio, Corrado and Berrocal, Veronica and
               Biggeri, Annibale},
  title     = {Preferential sampling and {B}ayesian geostatistics:
               Statistical modeling and examples},
  journal   = {Statistical Methods in Medical Research},
  year      = {2016},
  volume    = {25},
  number    = {4},
  pages     = {1224--1243},
  doi       = {10.1177/0962280216660409}
}

@article{GelfandShirota2019,
  author    = {Gelfand, Alan E. and Shirota, Shinichiro},
  title     = {Preferential sampling for presence/absence data and for
               fusion of presence/absence data with presence-only data},
  journal   = {Ecological Monographs},
  year      = {2019},
  volume    = {89},
  number    = {3},
  pages     = {e01372},
  doi       = {10.1002/ecm.1372}
}

@article{ShirotaGelfand2022,
  author    = {Shirota, Shinichiro and Gelfand, Alan E.},
  title     = {Preferential sampling for bivariate spatial data},
  journal   = {Spatial Statistics},
  year      = {2022},
  volume    = {51},
  pages     = {100674},
  doi       = {10.1016/j.spasta.2022.100674}
}

@article{Pennino2019,
  author    = {Pennino, Maria Grazia and Paradinas, Iosu and
               Illian, Janine B. and Mu\~{n}oz, Facundo and
               Bellido, Jos\'{e} Mar\'{\i}a and L\'{o}pez-Qu\'{\i}lez, Antonio
               and Conesa, David},
  title     = {Accounting for preferential sampling in species
               distribution models},
  journal   = {Ecology and Evolution},
  year      = {2019},
  volume    = {9},
  number    = {1},
  pages     = {653--663},
  doi       = {10.1002/ece3.4789}
}

@article{Conn2017,
  author    = {Conn, Paul B. and Thorson, James T. and Johnson, Devin S.},
  title     = {Confronting preferential sampling when analysing population
               distributions: diagnosis and model-based triage},
  journal   = {Methods in Ecology and Evolution},
  year      = {2017},
  volume    = {8},
  number    = {11},
  pages     = {1535--1546},
  doi       = {10.1111/2041-210X.12803}
}

@article{Lee2015,
  author    = {Lee, A. and Szpiro, A. and Kim, S.-Y. and Sheppard, L.},
  title     = {Impact of preferential sampling on exposure prediction and
               health effect inference in the context of air pollution
               epidemiology},
  journal   = {Environmetrics},
  year      = {2015},
  volume    = {26},
  number    = {4},
  pages     = {255--267},
  doi       = {10.1002/env.2334}
}

@article{Calhoun2024,
  author    = {Calhoun, Zachary D. and Black, Marilyn S. and
               Bergin, Mike and Carlson, David},
  title     = {Refining Citizen Climate Science: Addressing Preferential
               Sampling for Improved Estimates of Urban Heat},
  journal   = {Environmental Science \& Technology Letters},
  year      = {2024},
  volume    = {11},
  number    = {8},
  pages     = {845--850},
  doi       = {10.1021/acs.estlett.4c00296}
}

@article{Datta2016,
  author    = {Datta, Abhirup and Banerjee, Sudipto and
               Finley, Andrew O. and Gelfand, Alan E.},
  title     = {Hierarchical nearest-neighbor {G}aussian process models
               for large geostatistical datasets},
  journal   = {Journal of the American Statistical Association},
  year      = {2016},
  volume    = {111},
  number    = {514},
  pages     = {800--812},
  doi       = {10.1080/01621459.2015.1044091}
}

@article{Finley2019,
  author    = {Finley, Andrew O. and Datta, Abhirup and Cook, Bruce D.
               and Morton, Douglas C. and Andersen, Hans E. and
               Banerjee, Sudipto},
  title     = {Efficient algorithms for {B}ayesian nearest neighbor
               {G}aussian processes},
  journal   = {Journal of Computational and Graphical Statistics},
  year      = {2019},
  volume    = {28},
  number    = {2},
  pages     = {401--414},
  doi       = {10.1080/10618600.2018.1537924}
}

@book{Banerjee2015,
  author    = {Banerjee, Sudipto and Carlin, Bradley P. and
               Gelfand, Alan E.},
  title     = {Hierarchical Modeling and Analysis for Spatial Data},
  edition   = {2nd},
  publisher = {CRC Press},
  address   = {Boca Raton, FL},
  year      = {2014},
  doi       = {10.1201/b17115}
}

@article{AlbertChib1993,
  author    = {Albert, James H. and Chib, Siddhartha},
  title     = {Bayesian analysis of binary and polychotomous
               response data},
  journal   = {Journal of the American Statistical Association},
  year      = {1993},
  volume    = {88},
  number    = {422},
  pages     = {669--679},
  doi       = {10.1080/01621459.1993.10476321}
}

@article{Moller1998,
  author    = {M{\o}ller, Jesper and Syversveen, Anne Randi and
               Waagepetersen, Rasmus Plenge},
  title     = {Log {G}aussian {C}ox processes},
  journal   = {Scandinavian Journal of Statistics},
  year      = {1998},
  volume    = {25},
  number    = {3},
  pages     = {451--482},
  doi       = {10.1111/1467-9469.00115}
}

@article{Berman1992,
  author    = {Berman, Mark and Turner, T. Rolf},
  title     = {Approximating point process likelihoods with {GLIM}},
  journal   = {Applied Statistics},
  year      = {1992},
  volume    = {41},
  number    = {1},
  pages     = {31--38},
  doi       = {10.2307/2347614}
}

@article{EllicksonGrieco2013,
  author    = {Ellickson, Paul B. and Grieco, Paul L. E.},
  title     = {{Wal-Mart} and the geography of grocery retailing},
  journal   = {Journal of Urban Economics},
  volume    = {75},
  pages     = {1--14},
  year      = {2013},
  doi       = {10.1016/j.jue.2012.09.005}
}

@article{Karamshuk2013,
  author    = {Karamshuk, Dmytro and Noulas, Anastasios and
               Scellato, Salvatore and Nicosia, Vincenzo and
               Mascolo, Cecilia},
  title     = {Geo-{S}potting: Mining online location-based services
               for optimal retail store placement},
  journal   = {Proceedings of the 19th ACM SIGKDD International
               Conference on Knowledge Discovery and Data Mining},
  year      = {2013},
  pages     = {793--801},
  doi       = {10.1145/2487575.2487616}
}

@article{Kuo2002,
  author    = {Kuo, Ren-Jiun and Chi, Shie-Chin and Kao, Shih-Syun},
  title     = {A decision support system for selecting convenience
               store location through integration of fuzzy {AHP} and
               artificial neural network},
  journal   = {Computers in Industry},
  year      = {2002},
  volume    = {47},
  number    = {2},
  pages     = {199--214},
  doi       = {10.1016/S0166-3615(01)00147-6}
}

@article{Huff1963,
  author    = {Huff, David L.},
  title     = {A probabilistic analysis of shopping center trade areas},
  journal   = {Land Economics},
  year      = {1963},
  volume    = {39},
  number    = {1},
  pages     = {81--90},
  doi       = {10.2307/3144521}
}

@article{Applebaum1966,
  author    = {Applebaum, William},
  title     = {Methods for determining store trade areas, market
               penetration, and potential sales},
  journal   = {Journal of Marketing Research},
  year      = {1966},
  volume    = {3},
  number    = {2},
  pages     = {127--141},
  doi       = {10.1177/002224376600300202}
}

@article{Aversa2018,
  author    = {Aversa, Joseph and Doherty, Sean and Hernandez, Tony},
  title     = {Big data analytics: {T}he new boundaries of retail
               location decision making},
  journal   = {Papers in Applied Geography},
  year      = {2018},
  volume    = {4},
  number    = {4},
  pages     = {390--408},
  doi       = {10.1080/23754931.2018.1527720}
}

@article{Holmes2011,
  author    = {Holmes, Thomas J.},
  title     = {The diffusion of {W}al-{M}art and economies of density},
  journal   = {Econometrica},
  year      = {2011},
  volume    = {79},
  number    = {1},
  pages     = {253--302},
  doi       = {10.3982/ECTA7699}
}

@article{Jia2008,
  author    = {Jia, Panle},
  title     = {What happens when {W}al-{M}art comes to town:
               {A}n empirical analysis of the discount retailing
               industry},
  journal   = {Econometrica},
  year      = {2008},
  volume    = {76},
  number    = {6},
  pages     = {1263--1316},
  doi       = {10.3982/ECTA6649}
}

@techreport{VerPloeg2009,
  author    = {{Ver Ploeg}, Michele and Breneman, Vince and
               Farrigan, Tracey and Hamrick, Karen and Hopkins, David
               and Kaufman, Phil and Lin, Biing-Hwan and Nord, Mark
               and Smith, Travis A. and Williams, Ryan and
               Kinnison, Kelly and Olander, Carol and Singh, Anita and
               Tuckermanty, Elizabeth},
  title     = {Access to affordable and nutritious food: {M}easuring
               and understanding food deserts and their consequences:
               {R}eport to {C}ongress},
  institution = {U.S. Department of Agriculture, Economic Research
                 Service},
  year      = {2009},
  number    = {AP-036},
  address   = {Washington, DC}
}

@inproceedings{Davis2006,
  author    = {Davis, Jesse and Goadrich, Mark},
  title     = {The relationship between {P}recision-{R}ecall and
               {ROC} curves},
  booktitle = {Proceedings of the 23rd International Conference on
               Machine Learning},
  year      = {2006},
  pages     = {233--240},
  doi       = {10.1145/1143844.1143874}
}

@article{Saito2015,
  author    = {Saito, Takaya and Rehmsmeier, Marc},
  title     = {The precision-recall plot is more informative than
               the {ROC} plot when evaluating binary classifiers on
               imbalanced datasets},
  journal   = {PLOS ONE},
  year      = {2015},
  volume    = {10},
  number    = {3},
  pages     = {e0118432},
  doi       = {10.1371/journal.pone.0118432}
}

@article{Brier1950,
  author    = {Brier, Glenn W.},
  title     = {Verification of forecasts expressed in terms of probability},
  journal   = {Monthly Weather Review},
  year      = {1950},
  volume    = {78},
  number    = {1},
  pages     = {1--3},
  doi       = {10.1175/1520-0493(1950)078<0001:VOFEIT>2.0.CO;2}
}

%% ============================================================
%% APPENDICES
%% ============================================================
\appendix

\section{MCMC Convergence Diagnostics}
\label{app:convergence}

Table~\ref{tab:convergence} reports the effective sample size (ESS) for
key parameters of the PS Model. Inference is based on a single long chain, so
the Gelman--Rubin $\widehat{R}$ statistic is not available. We report ESS and
the marginal posterior histograms in Appendix~\ref{app:posteriors} as basic
diagnostics. The grid-level GP parameters and the loading $\delta$ have ESS
in the thousands, whereas the reported survival coefficients have much lower
ESS. These summaries do not replace a multi-chain convergence assessment.

\begin{table}[htbp]
  \centering
  \caption{MCMC effective sample sizes (ESS) for key parameters of the
           PS Model, based on the $3{,}000$ retained posterior draws.}
  \label{tab:convergence}
  \begin{tabular}{lc}
    \toprule
    Parameter & ESS \\
    \midrule
    $\delta$   & 2077 \\
    $\alpha_0$ & 431 \\
    $\sigma^2_*$ (grid GP) & 1835 \\
    $\phi_*$ (grid GP)     & 2825 \\
    $\beta$ (Nearest Supermarket Dist.) & 114 \\
    $\beta$ (Stay\_Home)  & 190 \\
    \bottomrule
  \end{tabular}
\end{table}

\section{Posterior Distributions of Regression Coefficients and Hyperparameters}
\label{app:posteriors}

\begin{figure}[htbp]
  \centering
  \includegraphics[width=\textwidth]{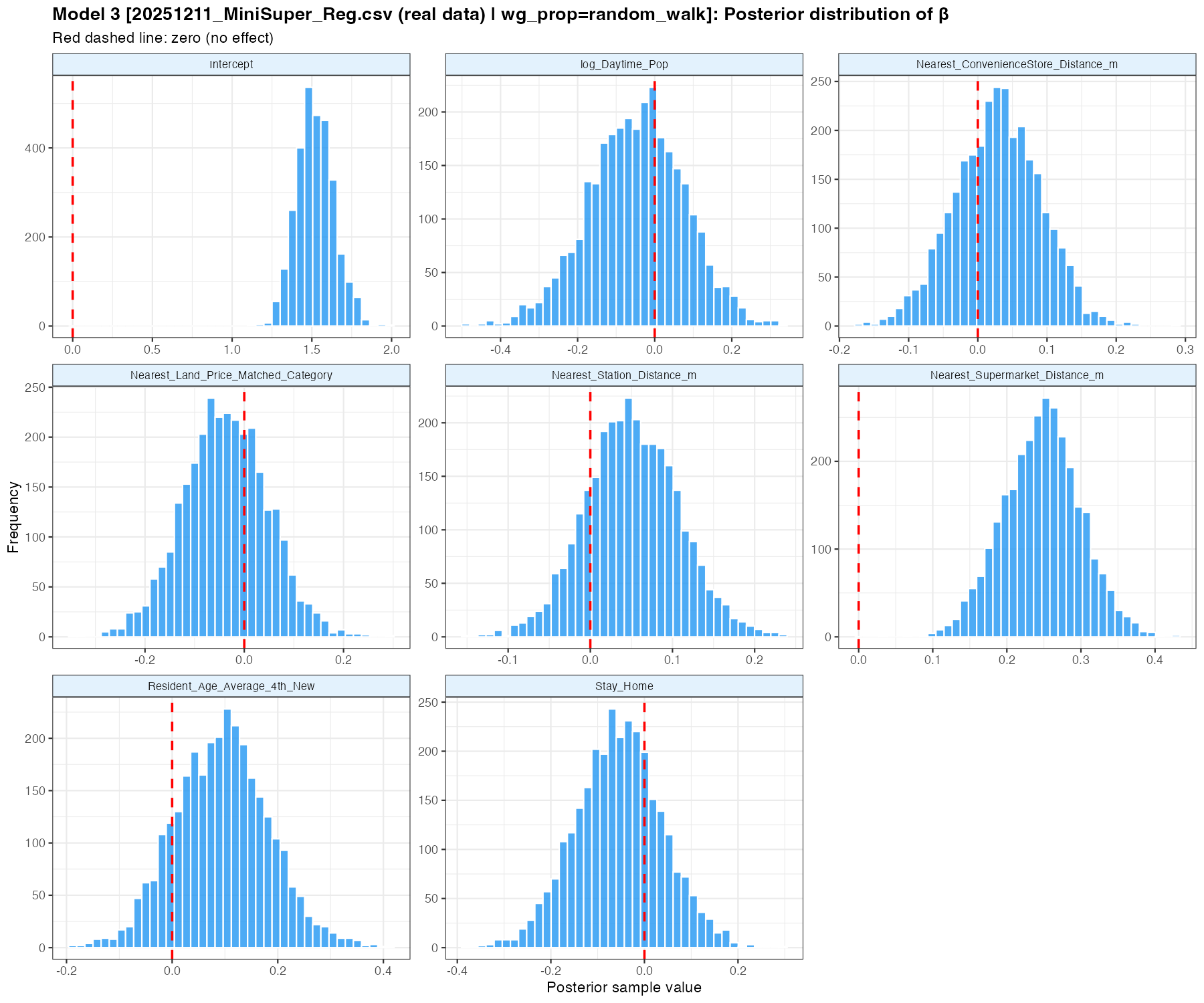}
  \caption{Posterior histograms for the probit regression
           coefficients $\bbeta$, including the intercept (PS Model).
           The red dashed line marks zero. Histograms use the
           $3{,}000$ retained draws (first $30{,}000$ iterations
           discarded as burn-in, thinned by $10$).}
  \label{fig:trace_beta}
\end{figure}

\begin{figure}[htbp]
  \centering
  \includegraphics[width=0.85\textwidth]{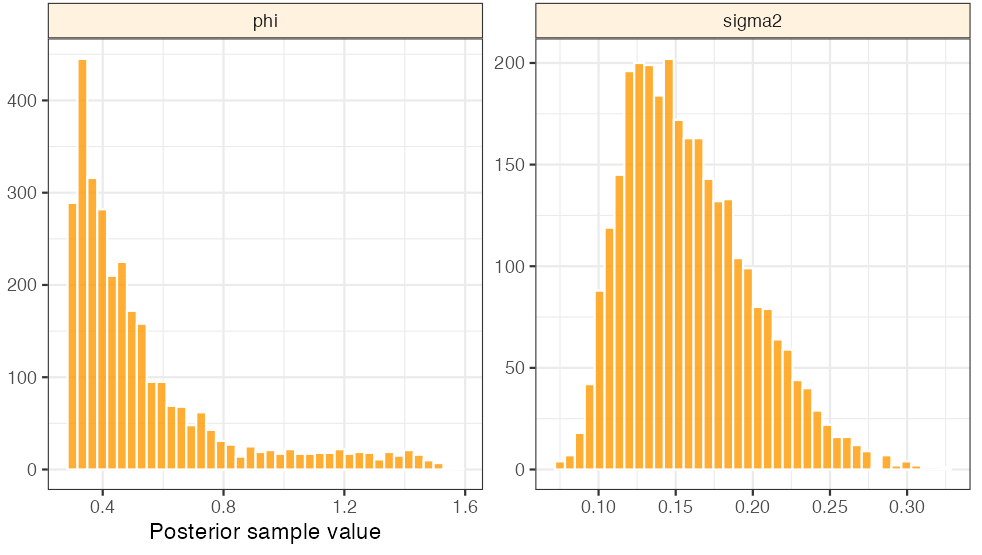}
  \caption{Posterior histograms for the grid-GP hyperparameters
           $\sigma^2_*$ and $\phi_*$ (PS Model).}
  \label{fig:trace_hyper}
\end{figure}

Figures~\ref{fig:trace_beta}--\ref{fig:trace_hyper} show posterior
histograms for all $\bbeta$ coefficients (including the probit intercept)
and for the grid-GP hyperparameters $(\sigma^2_*, \phi_*)$, from the PS Model.

\section{LGCP Covariate Coefficients}
\label{app:gamma}

Table~\ref{tab:gamma} reports posterior summaries for the LGCP
covariate coefficients $\bgamma$, which describe the association
between neighborhood characteristics and the \textit{location} of
mini-supermarkets (as opposed to their survival).

\begin{table}[htbp]
  \centering
  \caption{Posterior summaries for LGCP covariate coefficients
           $\bgamma$ (standalone LGCP model). The PS Model and Model~4 yield
           nearly identical estimates (each coefficient differs by at most
           $\approx 0.02$). Positive values indicate that higher values of the
           covariate are associated with higher local store density.}
  \label{tab:gamma}
  \begin{tabular}{lrrrr}
    \toprule
    Variable & Mean & SD & 2.5\% & 97.5\% \\
    \midrule
    Nearest\_Supermarket\_Distance\_m      & $\phantom{-}0.010$ & 0.035 & $-0.057$ & $\phantom{-}0.078$ \\
    Nearest\_ConvenienceStore\_Distance\_m & $-0.192$ & 0.037 & $-0.263$ & $-0.119$ \\
    Nearest\_Station\_Distance\_m          & $-0.339$ & 0.036 & $-0.406$ & $-0.265$ \\
    Stay\_Home                             & $\phantom{-}0.181$ & 0.052 & $\phantom{-}0.080$ & $\phantom{-}0.283$ \\
    Resident\_Age\_Average\_4th\_New       & $-0.174$ & 0.042 & $-0.258$ & $-0.090$ \\
    Daytime\_Pop                           & $\phantom{-}0.629$ & 0.067 & $\phantom{-}0.497$ & $\phantom{-}0.761$ \\
    Nearest\_Land\_Price & $-0.056$ & 0.055 & $-0.161$ & $\phantom{-}0.054$ \\
    \midrule
    $\alpha_0$ (LGCP intercept)            & $\phantom{-}6.463$ & 0.048 & $\phantom{-}6.369$ & $\phantom{-}6.555$ \\
    \bottomrule
  \end{tabular}
\end{table}

\section{Other Figures and Tables}
\label{app:figures}

\begin{figure}[htbp]
  \centering
  \includegraphics[width=\textwidth]{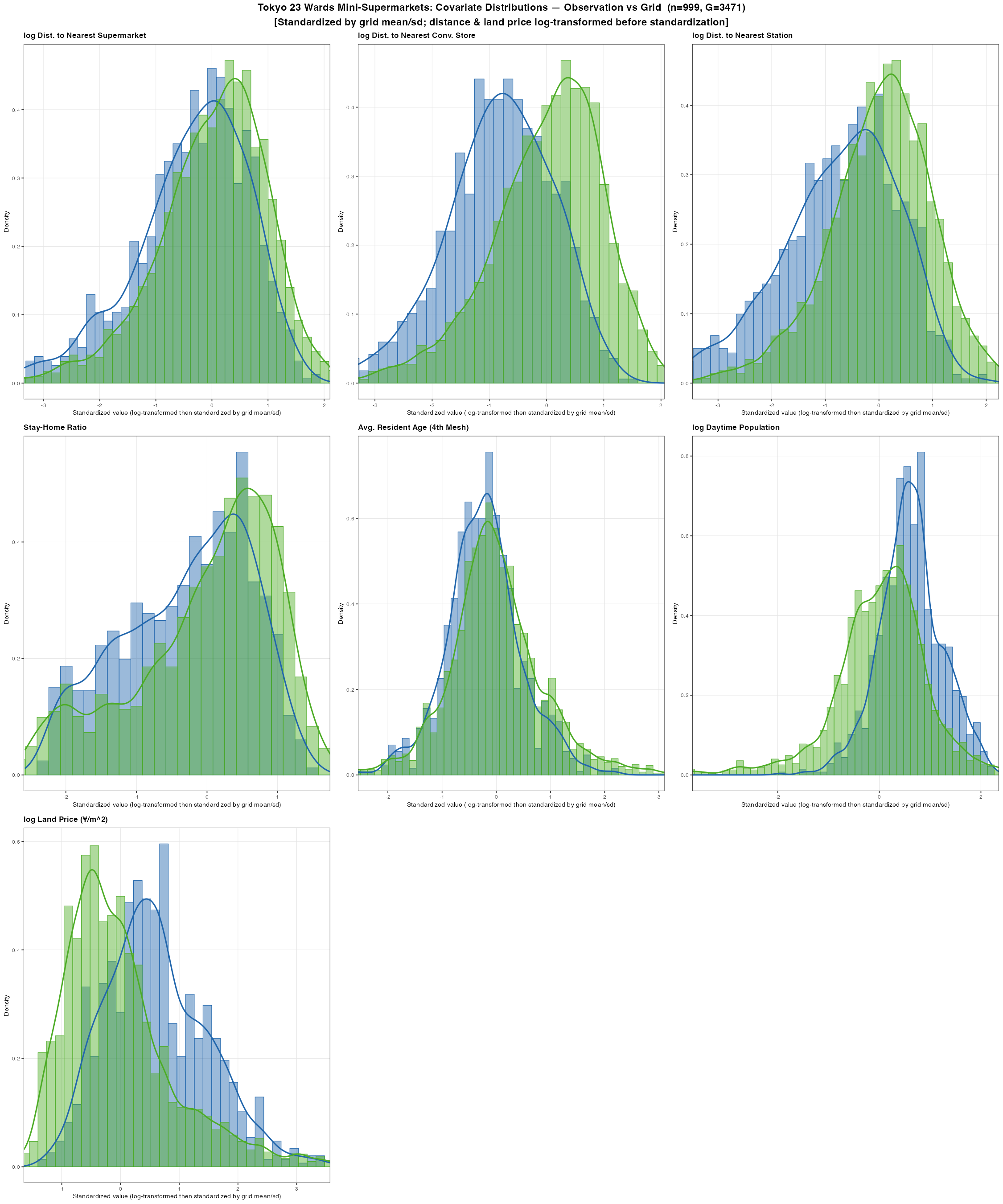}
  \caption{Covariates by mini-supermarket locations vs grids}
  \label{fig:hist_obs_grid}
\end{figure}

\begin{figure}[htbp]
  \centering
  \includegraphics[width=\textwidth]{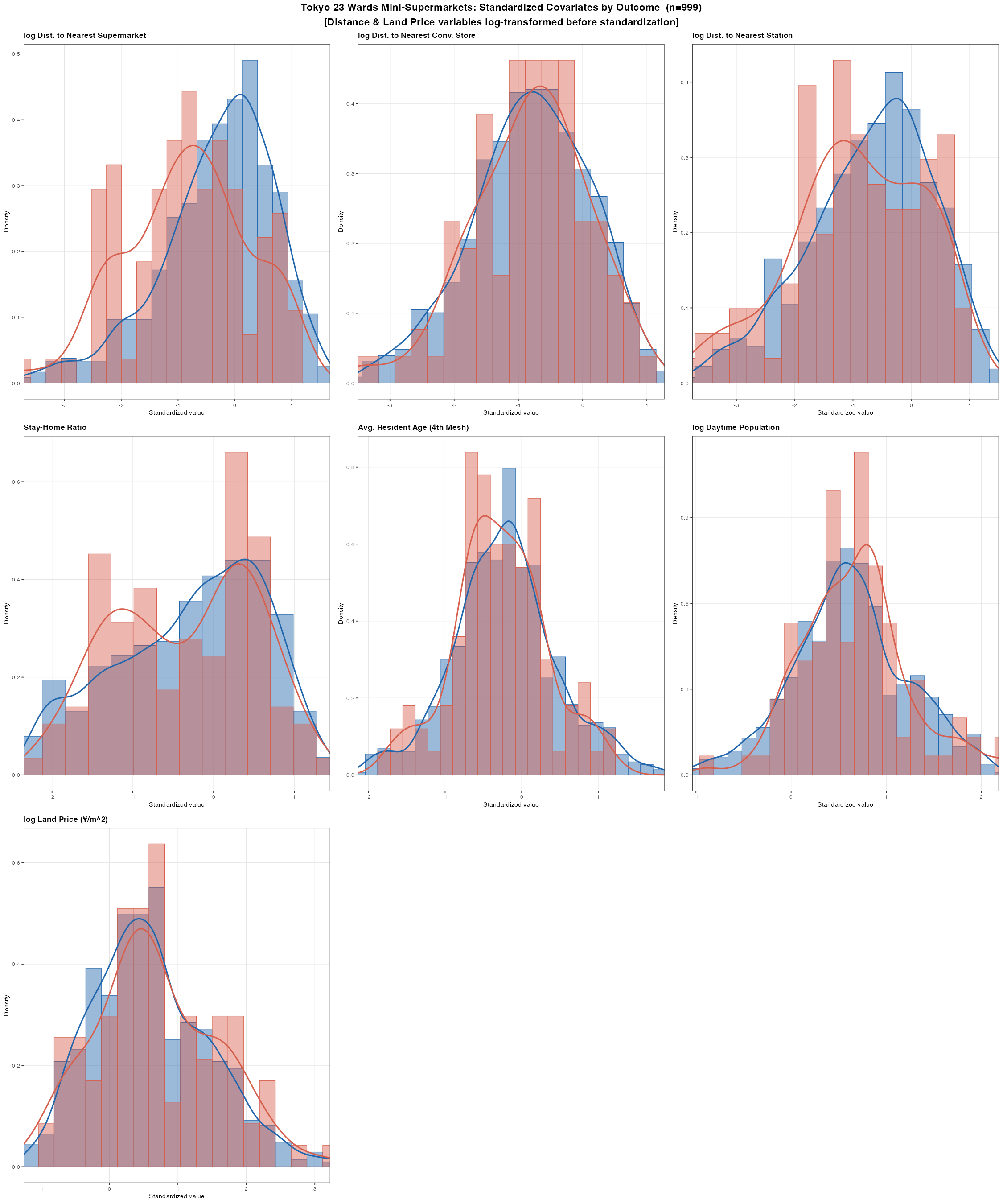}
  \caption{Covariates by outcomes.}
  \label{fig:hist_by_y}
\end{figure}

\end{document}